\documentclass[english,aps, onecolumn,notitlepage, pre,superscriptaddress, notitlepage, longbibliography , dvipsnames]{revtex4-1}
\usepackage[T1]{fontenc}
\usepackage{array}
\usepackage{booktabs}
\usepackage{amsmath}

\makeatletter

\providecommand{\tabularnewline}{\\}

\usepackage{amsmath}
\usepackage{subfigure}
\usepackage{graphicx, mathtools}
\usepackage[normalem]{ulem}
\usepackage[dvipsnames]{xcolor}
\usepackage[colorlinks=true,linkcolor=MidnightBlue,urlcolor=MidnightBlue,citecolor=MidnightBlue,anchorcolor=MidnightBlue]{hyperref}
\usepackage[T1]{fontenc}
\usepackage[utf8]{inputenc}

\makeatother

\usepackage{babel}
\begin{document}
\title{Stokes traction on an active particle}
\author{G\"{u}nther Turk}
\email{gt369@cam.ac.uk}

\affiliation{DAMTP, Centre for Mathematical Sciences, University of Cambridge,
Wilberforce Road, Cambridge CB3 0WA, United Kingdom}
\author{Rajesh Singh}
\affiliation{DAMTP, Centre for Mathematical Sciences, University of Cambridge,
Wilberforce Road, Cambridge CB3 0WA, United Kingdom}
\affiliation{Department of Physics, IIT Madras, Chennai 600036, India}
\author{Ronojoy Adhikari}
\affiliation{DAMTP, Centre for Mathematical Sciences, University of Cambridge,
Wilberforce Road, Cambridge CB3 0WA, United Kingdom}
\affiliation{\textsuperscript{}The Institute of Mathematical Sciences-HBNI, CIT
Campus, Chennai 600113, India }
\begin{abstract}
The mechanics and statistical mechanics of a suspension of active
particles are determined by the traction (force per unit area) on
their surfaces. Here we present an exact solution of the direct boundary
integral equation for the traction on a spherical active particle
in an imposed slow viscous flow. Both single- and double-layer integral
operators can be simultaneously diagonalised in a basis of irreducible
tensorial spherical harmonics and the solution, thus, can be presented
as an infinite number of linear relations between the harmonic coefficients
of the traction and the velocity at the boundary of the particle.
These generalise Stokes laws for the force and torque. Using these
relations we obtain simple expressions for physically relevant quantities
such as the symmetric-irreducible dipole acting on, or the power dissipated
by, an active particle in an arbitrary imposed flow. We further present
an explicit expression for the variance of the Brownian contributions
to the traction on an active colloid in a thermally fluctuating fluid.
\end{abstract}
\maketitle

\section{Introduction}

A passive colloidal particle produces flow in the ambient fluid when
it translates or rotates. In contrast, an active particle can produce
a flow even when stationary \citep{paxton2004,howse2007self,jiang2010active,ebbens2010pursuit,palacci2013living}.
Examples include microorganisms \citep{brennen1977} and autophoretic
particles \citep{ebbens2010pursuit}. The exterior flow of active
particles is due to local non-equilibrium processes such as ciliary
motion (in the case of microorganisms) and osmotic flows (in the case
of autophoretic particles). These non-equilibrium processes, when
confined to a thin layer at the surface of the particles, can be modelled
by adding a surface slip $\text{\ensuremath{\boldsymbol{v}}}^{\mathcal{A}}$
to the commonly used no-slip boundary condition on particle surfaces
\citep{lighthill1952,blake1971a,anderson1989colloid}. The surface
slip sets the ambient fluid in motion, causing stresses that react
back on the particle. For a rigid particle, these integrate to a net
force and a net torque on the particle centre of mass. Since fluid
inertia is negligible at the colloidal scale, fluid motion is governed
by the Stokes equation. The solution of the Stokes equation with prescribed
velocity boundary conditions provides the stress in the fluid and,
when evaluated on the particle, the traction (force per unit area)
\citep{stokesEffectInternalFriction1850,landau1959fluid,brunn1976effect,brunnMotionSlightlyDeformed1979,pak2014generalized,pedley2016spherical,pedley_brumley_goldstein_2016,rojas2021hydrochemical}.
The boundary integral formulation of the Stokes equation provides
an alternative route to obtaining the traction that obviates the need
to solve for the fluid flow in the bulk \citep{lorentz1907,fkg1930bandwertaufgaben,ladyzhenskaya1969,cheng2005heritage}.
Instead, it provides a direct linear integral relation between quantities
that are defined only at the boundaries, namely the traction and the
velocity boundary condition. The boundary integral formulation has
been used extensively to describe the dynamics of passive colloidal
particles \citep{youngren1975stokes,felderhof1976,felderhofForceDensityInduced1976a,zick1982stokes,schmitzCreepingFlowSpherical1982}
and, more recently, of active colloidal particles \citep{ghose2014irreducible,singh2015many,singh2017fluctuation,singh2018generalized}. 

Despite the large body of work on the integral equation approach to
active particle dynamics, the simplest problem of a single active
sphere in an unbounded fluid has not been solved exactly. Apart from
its intrinsic theoretical interest, such a solution is of potential
use in numerical solutions of the boundary integral equation for many
particles, where numerical iterations can be initialised with the
exact one-particle solution. It is known that discretisations of boundary
integral equations for this class of problems leads to diagonally
dominant linear systems and the one-particle solution is the exact
solution when hydrodynamic interactions are ignored. This suggests
that iterations initialised at the one-particle solution can converge
rapidly to the diagonally-dominant numerical solutions \citep{singh2018generalized}.

In this paper we solve the direct boundary integral equation exactly
for the traction on a spherical active particle in an unbounded fluid.
Expansion in a complete basis followed by the minimisation of the
residual is a convenient strategy for solving linear integral equations.
In this so-called Ritz-Galerkin procedure \citep{boyd2001chebyshev,finlayson1966method,singh2015many},
a basis that yields a diagonal linear system is particularly useful
as the system, then, is trivially soluble. The direct boundary integral
equation contains a pair of integral operators -- the single-layer
and double-layer operators -- and it is not obvious that a basis
that diagonalises one operator will necessarily diagonalise the other.
Here we show that the basis of tensorial spherical harmonics (TSH)
simultaneously diagonalises both the single-layer and double-layer
integral operators and, in this sense, provides the most appropriate
choice of basis. The boundary integral equation is reduced, thereby,
to an infinite-dimensional diagonal linear system that can be solved
trivially. We obtain compact, closed-form linear relations between
the harmonic modes of the traction and the boundary velocity. The
first two of these are the familiar Stokes laws for the force and
torque of a spherical particle of radius $b$ in rigid body motion
in an unbounded fluid of dynamic viscosity $\eta$ containing the
scalar friction coefficients $6\pi\eta b$ and $8\pi\eta b^{3}$,
respectively \citep{stokesEffectInternalFriction1850,happel1965low}. 

In what follows, we present our solution and some implications thereof
in detail. In Section \ref{sec:Main-results} we discuss our main
findings -- exact linear relations between the traction and the boundary
velocity on an active particle in an imposed flow in terms of scalar
generalised friction coefficients. We refer to these relations as
generalised Stokes laws and emphasise that the friction coefficients
due to imposed flow and activity are distinct. We then turn towards
their derivation in Section \ref{sec:Derivation}. We briefly recall
the boundary integral representation of Stokes flow for three distinct
contributions to the traction on the surface of an active particle
in an imposed flow. These are (a) rigid body motion, (b) imposed flow,
and (c) active slip. Using spectral expansions and Ritz-Galerkin discretisation
of the boundary integral equations we derive an exact solution thereof
in terms of matrix elements of the single- and double-layer integrals.
These matrix elements are found to diagonalise simultaneously in a
basis of TSH. The resulting linear system of equations is thus solved
trivially to find the generalised Stokes laws. In Section \ref{sec:stress}
we discuss a number of applications of our findings. First, we derive
an expression relating the expansion coefficients of the imposed flow,
expanded in TSH on the surface of the sphere, with its Taylor expansion
about the centre of the particle, and thus relate our work to the
generalised Fax{\'e}n  relations \citep{brunn1980faxen}. We then
discuss the symmetric-irreducible dipole on an active particle in
an imposed straining flow. In terms of the previously derived friction
coefficients we then obtain an expression for the power dissipated
by an active particle. Finally, we present an explicit expression
for the variance of the Brownian contributions to the traction on
an active colloid in a thermally fluctuating system. We conclude in
Section \ref{sec:Discussion} by summarising our results, putting
them into context with previous work, and suggesting directions for
future research. 

\section{results\label{sec:Main-results}}

In this section we briefly outline our main results for the traction
on an active colloidal particle due to the most general form of surface
velocity and arbitrary imposed flow $\boldsymbol{v}^{\infty}(\boldsymbol{r})$.
We consider a spherical active particle of radius $b$ in an incompressible
fluid of viscosity $\eta$. The boundary condition at the surface
of the particle is

\begin{equation}
\boldsymbol{v}(\boldsymbol{R}+\boldsymbol{\rho})=\boldsymbol{V}+\boldsymbol{\Omega}\times\boldsymbol{\rho}+\boldsymbol{v}^{\mathcal{A}}(\boldsymbol{\rho})=\boldsymbol{v}^{\mathcal{D}}(\boldsymbol{\rho})+\boldsymbol{v}^{\mathcal{A}}(\boldsymbol{\rho}).\label{eq:boundary-1}
\end{equation}
The rigid body motion \textbf{$\boldsymbol{v}^{\mathcal{D}}$} is
specified by the translational velocity $\boldsymbol{V}$ and angular
velocity $\mathbf{\boldsymbol{\Omega}}$ of the particle. Here, $\boldsymbol{R}$
is the centre of the colloid, $\boldsymbol{\rho}$ is its radius vector
and $\boldsymbol{v}^{\mathcal{A}}$ is its active slip velocity. The
only restriction on the active slip is that it conserves mass in the
fluid, ie
\begin{equation}
\int\hat{\boldsymbol{\rho}}\cdot\boldsymbol{v}^{\mathcal{A}}\,d\mathcal{S}=0,
\end{equation}
where $\mathcal{S}$ is the surface of the colloid and $\hat{\boldsymbol{\rho}}$
is the unit normal vector to the surface of the colloid, pointing
into the surrounding fluid.

It is convenient to express the traction on the particle as a sum
of three distinct contributions

\begin{equation}
\boldsymbol{f}=\boldsymbol{f}^{\mathcal{D}}+\boldsymbol{f}^{\infty}+\boldsymbol{f}^{\mathcal{A}}.\label{eq:traction}
\end{equation}
Here, $\boldsymbol{f}^{\mathcal{D}}$ is the traction due to the colloid's
rigid body motion $\boldsymbol{v}^{\mathcal{D}}$ alone, $\boldsymbol{f}^{\infty}$
represents the traction on a no-slip particle when held stationary
in an imposed flow $\boldsymbol{v}^{\infty}$, and $\boldsymbol{f}^{\mathcal{A}}$
is the contribution from active surface slip $\boldsymbol{v}^{\mathcal{A}}$,
see Appendix \ref{app:Derivation-boundary-integral}.

In order to parametrise the surface fields on the boundary of the
active particle, we expand the velocity and the traction at the colloid's
surface in tensorial spherical harmonics (TSH) $\boldsymbol{Y}^{(l)}(\hat{\boldsymbol{\rho}})$
as

\begin{gather}
\boldsymbol{v}^{\lambda}(\boldsymbol{R}+\boldsymbol{\rho})=\sum_{l=1}^{\infty}w_{l}\boldsymbol{V}^{\lambda(l)}\odot\boldsymbol{Y}^{(l-1)}(\hat{\boldsymbol{\rho}}),\quad\boldsymbol{f}^{\lambda}(\boldsymbol{R}+\boldsymbol{\rho})=\sum_{l=1}^{\infty}\tilde{w}_{l}\boldsymbol{F}^{\lambda(l)}\odot\boldsymbol{Y}^{(l-1)}(\hat{\boldsymbol{\rho}}),\label{eq:expansion-v}
\end{gather}
where $\lambda\in\left\{ \mathcal{D},\infty,\mathcal{A}\right\} $,
and 
\begin{equation}
w_{l}=\frac{1}{(l-1)!(2l-3)!!},\qquad\tilde{w}_{l}=\frac{2l-1}{4\pi b^{2}}.
\end{equation}
The product $\odot$ represents a maximal contraction of indices between
two tensors. The TSH are defined as
\begin{equation}
Y_{\alpha_{1}\dots\alpha_{l}}^{(l)}(\hat{\boldsymbol{\rho}})=(2l-1)!!\Delta_{\alpha_{1}\dots\alpha_{l},\beta_{1}\dots\beta_{l}}^{(l)}\hat{\rho}_{\beta_{1}}\dots\hat{\rho}_{\beta_{l}}=(-1)^{l}\,\rho^{l+1}\,\nabla_{\alpha_{1}}\dots\nabla_{\alpha_{l}}\frac{1}{\rho},
\end{equation}
with $\rho=\left\Vert \boldsymbol{\rho}\right\Vert _{2}$, where $\left\Vert \cdot\right\Vert _{2}$
is the Euclidean norm and $\boldsymbol{\Delta}{}^{(l)}$ is a rank
$2l$ tensor, which projects a tensor of rank $l$ onto its symmetric
and traceless part. Excellent summaries of their properties and the
identities they obey are available in the literature \citep{brunn1976effect,brunnMotionSlightlyDeformed1979,hess2015tensors}.

By definition, $\boldsymbol{F}^{\lambda(l)}$ and $\boldsymbol{V}^{\lambda(l)}$
are symmetric-irreducible in their last $l-1$ indices, and thus can
each be expressed as the sum of three irreducible tensors, $\boldsymbol{F}^{\lambda(l\sigma)}$
and $\boldsymbol{V}^{\lambda(l\sigma)}$, with the index $\sigma\in\{s,a,t\}$
labelling the symmetric-irreducible (rank $l$), the antisymmetric
(rank $l-1$), and the trace (rank $l-2$) parts of the reducible
tensors, respectively \citep{hess2015tensors}. This decomposition
and the projection of the expansion coefficients onto their irreducible
subspaces are given by
\begin{equation}
\boldsymbol{F}^{\lambda(l)}=\boldsymbol{D}^{(l\sigma)}\odot\boldsymbol{F}^{\lambda(l\sigma)},\qquad\boldsymbol{F}^{\lambda(l\sigma)}=\boldsymbol{P}^{(l\sigma)}\odot\boldsymbol{F}^{\lambda(l)},\label{eq:decomp-proj}
\end{equation}
respectively, with analogous expressions for the velocity coefficients.
Repeated mode indices $(l\sigma)$ are summed over implicitly for
the decomposition operators $\boldsymbol{D}^{(l\sigma)}$. Both the
decomposition operators and the projection operators are explicitly
defined in Section \ref{sec:Stokes-traction}.

As derived in Sections \ref{sec:frictionTensors} and \ref{sec:Stokes-traction},
the generalised Stokes laws for an isolated active particle in an
unbounded domain are

\begin{equation}
\boldsymbol{F}^{\mathcal{D}(l\sigma)}=-\gamma_{l\sigma}\,\boldsymbol{V}^{\mathcal{D}(l\sigma)},\quad\boldsymbol{F}^{\infty(l\sigma)}=\gamma_{l\sigma}\,\boldsymbol{V}^{\infty(l\sigma)},\quad\boldsymbol{F}^{\mathcal{A}(l\sigma)}=-\hat{\gamma}_{l\sigma}\,\boldsymbol{V}^{\mathcal{A}(l\sigma)},\label{eq:scalarStokes}
\end{equation}
for which we can give the \emph{scalar }generalised friction coefficients
exactly to \emph{arbitrary} order in $l$ as
\begin{align}
 & \gamma_{ls}=\frac{4\pi\eta b\,\left(2l+1\right)}{\left(l+1\right)\left(l-1\right)!\left(2l-3\right)!!}, &  & \gamma_{la}=\frac{4\pi\eta b}{\left(l-1\right)!\left(2l-3\right)!!}, &  & \gamma_{lt}=\frac{4\pi\eta b}{(l-2)\left(l-1\right)!\left(2l-5\right)!!},\nonumber \\
 & \hat{\gamma}_{ls}=\frac{4\pi\eta b\,\left(2l^{2}+1\right)}{\left(l+1\right)\left(l-1\right)!\left(2l-1\right)!!}, &  & \hat{\gamma}_{la}=\frac{4\pi\eta b\,\left(l+1\right)}{\left(l-1\right)!\left(2l-1\right)!!}, &  & \hat{\gamma}_{lt}=\frac{8\pi\eta b\,l}{\left(l-1\right)!\left(2l-1\right)!!}.\label{eq:scalarFriction}
\end{align}
 It should be noted that while the friction coefficients due to
imposed fluid flow ($\gamma_{l\sigma}$) and those due to active surface
slip ($\hat{\gamma}_{l\sigma}$) are equivalent for the modes of rigid
body motion, see Section \ref{sec:stress}, they are in general distinct.
The difference is due to the double-layer integral in the boundary
integral equations \eqref{eq:BIE-3parts}. The friction coefficients
$\gamma_{1s}$, $\gamma_{2s}$, and $\gamma_{2a}$ are available in
the literature in terms of a Taylor expansion of the imposed flow
about the centre of the particle and referred to as the Fax{\'e}n 
relations \citep{faxen1922widerstand,batchelorGreen1972,rallison1978note}.
On the other hand, our results have been obtained in terms of expansion
coefficients of the imposed flow for arbitrary $(l\sigma)$. We derive
a relation between these two approaches in Section \ref{subsec:Expansion-coefficients-vs}.
More generally, analogous expressions to $\gamma_{l\sigma}$ for arbitrary
modes $(l\sigma)$ have been obtained by various authors \citep{brunn1976effect,brunnMotionSlightlyDeformed1979,felderhof1976,felderhofForceDensityInduced1976a,schmitzCreepingFlowSpherical1982};
see Table \ref{tab:history}.
\begin{figure}
\centering 
\includegraphics[width=0.94\columnwidth]{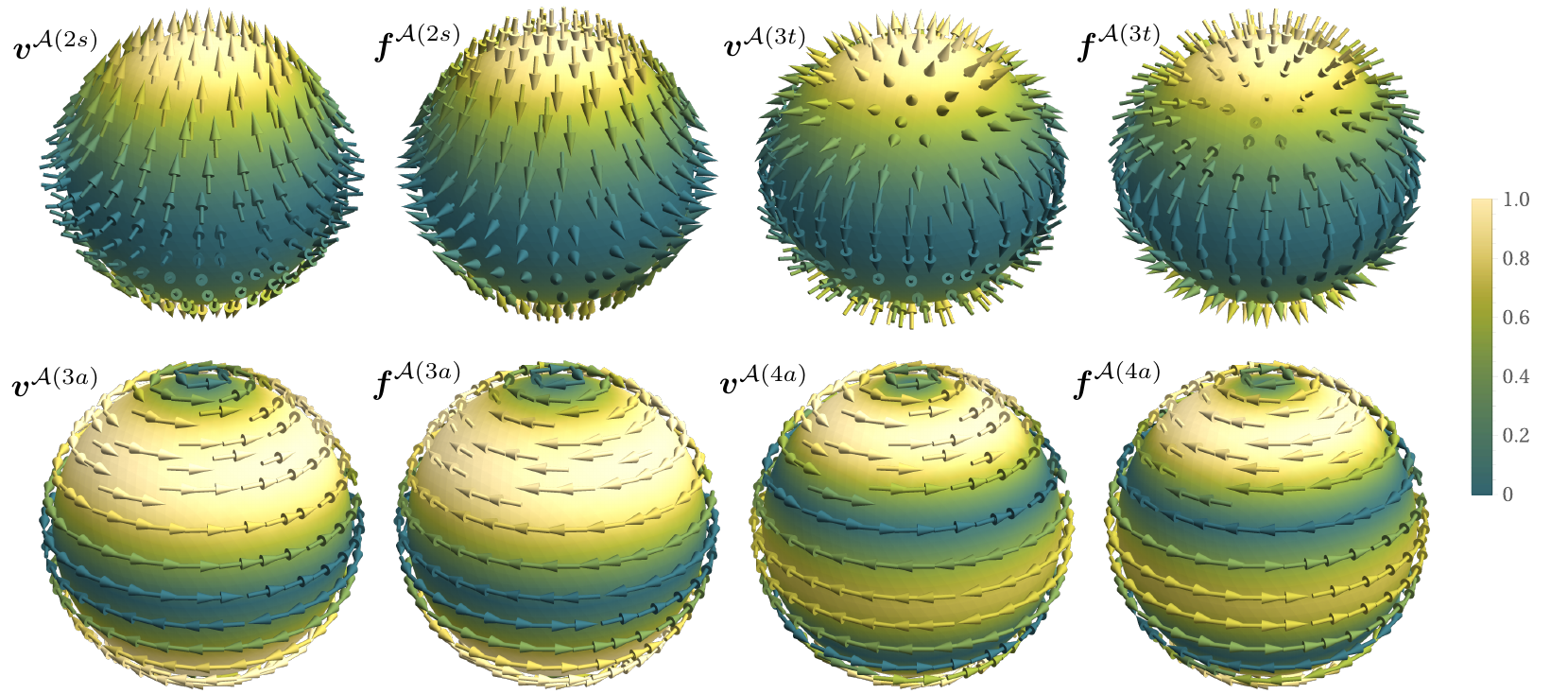}

\caption{\label{fig:Slip-and-traction}Panels denoted by $\boldsymbol{v}^{\mathcal{A}(l\sigma)}$
($\boldsymbol{f}^{\mathcal{A}(l\sigma)}$) show the vector field plots
of the slip (traction) due to an isolated $(l\sigma)$ mode of the
expansion \eqref{eq:expansion-v}. Here, the irreducible tensors $\boldsymbol{V}^{\lambda(l\sigma)}$
are naturally parametrised in terms of the TSH as follows: $\mathbf{V}^{\lambda(ls)}=V_{ls}^{0,\lambda}\,\mathbf{Y}^{(l)}(\boldsymbol{e}),$
$\mathbf{V}^{\lambda(la)}=V_{la}^{0,\lambda}\,\mathbf{Y}^{(l-1)}(\boldsymbol{e}),$
and $\mathbf{V}^{\lambda(lt)}=V_{lt}^{0,\lambda}\,\mathbf{Y}^{(l-2)}(\boldsymbol{e}),$
where the \emph{uniaxial} parameterisations are defined in terms of
the orientation vector $\boldsymbol{e}$ of the active particle and
$V_{l\sigma}^{0,\lambda}$ are the scalar strengths of the modes.
From this parametrisation, it follows that the $\mathbf{V}^{\lambda(l\sigma)}$
are either even (apolar) or odd (polar) under inversion symmetry $\boldsymbol{e}\rightarrow-\boldsymbol{e}$
with respect to the orientation of the particle. The figure shows
the vector fields $\boldsymbol{v}^{\mathcal{A}}$ and $\boldsymbol{f}^{\mathcal{A}}$
due to the leading modes of apolar $(2s)$, polar $(3t)$, achiral
$(3a)$, and chiral $(4a)$ symmetry. The fields have been plotted
on the surface of the particle with orientation $\boldsymbol{e}$
along the north pole. For clarity, we have lifted the vector field
off the surface slightly, while its magnitude has been overlaid on
the surface. It follows directly from the \emph{scalar} friction coefficients
of the generalised Stokes laws \eqref{eq:scalarStokes} that both
slip and traction exhibit the same symmetry.}
\end{figure}
\begin{table*}
\renewcommand{\arraystretch}{1.6} \centering

\begin{tabular*}{1\textwidth}{@{\extracolsep{\fill}}>{\raggedright}p{3.1cm}>{\raggedright}p{3.4cm}>{\raggedright}p{3.4cm}>{\raggedright}p{6.4cm}}
\toprule 
 & Boundary condition & Expansion basis & Methodology\tabularnewline
\midrule
Stokes \citep{stokesEffectInternalFriction1850} & No-slip (passive) &  & \tabularnewline
Lighthill \citep{lighthill1952} and Blake \citep{blake1971a} & Axisymmetric slip (active) & Scalar harmonics & Lamb's general solution and scalar harmonic expansion of boundary
condition to obtain coefficients of the expansion\tabularnewline
Felderhof and Schmitz \citep{felderhof1976,felderhofForceDensityInduced1976a,schmitzCreepingFlowSpherical1982} & Mixed slip-stick (passive) in imposed flow & Vector spherical harmonics (VSH) & BIE for a passive no-slip sphere. Single-layer diagonalises under
VSH (Antenna theorems), obtained $\gamma$\tabularnewline
Brunn \citep{brunn1976effect,brunnMotionSlightlyDeformed1979} & Mixed slip-stick (passive) in imposed flow & Tensor spherical harmonics (TSH) & Lamb's general solution in terms of multipole potentials, using boundary
conditions to find coefficients, obtained $\gamma$\tabularnewline
Ghose and Adhikari \citep{ghose2014irreducible} & General slip & TSH & Indirect formulation of the BIE. Friction tensors for the first few
modes.\tabularnewline
Pak and Lauga \citep{pak2014generalized}, Pedley et al \citep{pedley2016spherical,pedley_brumley_goldstein_2016} & General slip & Scalar harmonics & Extend Lighthill and Blake's calculation to include azimuthal slip
using Lamb's general solution\tabularnewline
This paper & General slip & TSH & Direct formulation of the BIE, introduced in \citep{singh2015many,singh2018generalized}.
Friction tensors for all modes due to slip and imposed flow.\tabularnewline
\bottomrule
\end{tabular*}\caption{\label{tab:history}Chronology of analytical results for the traction
on a single spherical particle in an unbounded Stokes flow. Here ``active''
implies a sphere with active surface slip, while ``passive'' implies
a sphere with no-slip, or alternatively a slip-stick boundary condition.
In the latter, passive case, slip must be interpreted as a passive
feature, comparable to the slippage at a boundary of the fluid domain
\citep{barratInfluenceWettingProperties1999,laugaBrownianMotionPartialslip2005,ketzetziSlipLengthDependent2020}.
The two main approaches to obtain higher order friction coefficients
are (a) using Lamb's general solution to obtain the flow field around
a particle, from which the stress tensor and thus the traction can
be derived, and (b) solving the boundary integral equation (BIE) to
obtain the traction directly.}
\end{table*}

It is intuitive that in the unbounded domain the generalised Stokes
laws, expressing the linear relations between the irreducible modes
of the traction and the corresponding modes of the boundary velocity,
must be scalar relations due to symmetry considerations. A visualisation
of this in terms of the active slip velocity and the resulting hydrodynamic
traction due to a given ($l\sigma$) mode is shown in Figure \ref{fig:Slip-and-traction}. 

\section{Derivation\label{sec:Derivation}}

This section is dedicated to the derivation of the generalised Stokes
laws, Eq. \eqref{eq:scalarStokes}. We revisit the boundary integral
formulation of the Stokes equation and define the linearly independent
boundary integral equations for the contributions to the force per
unit area (traction) on the particle due to (a) rigid body motion,
(b) imposed flow, and (c) active slip. We then solve these boundary
integral equations exactly, using spectral expansions and Ritz-Galerkin
discretisation. The matrix elements of the resulting linear system
of equations are solved for in Fourier space, and found to diagonalise
simultaneously in a basis of tensorial spherical harmonics. This diagonalisation
results directly in the generalised Stokes laws.

\subsection{Boundary integral formulation of the Stokes equation\label{sec:biStokes}}

We recall the boundary integral equation for a particle with boundary
conditions given by Eq. \eqref{eq:boundary-1} in an imposed flow
$\boldsymbol{v}^{\infty}(\boldsymbol{r})$. Incompressibility of the
fluid implies $\boldsymbol{\nabla}\cdot\boldsymbol{v}=0$. At the
colloidal scale, the fluid satisfies the Stokes equation, $\boldsymbol{\nabla}\cdot\boldsymbol{\sigma}=0$,
with the Cauchy stress tensor $\sigma_{\alpha\beta}=-p\delta_{\alpha\beta}+\eta\left(\nabla_{\alpha}v_{\beta}+\nabla_{\beta}v_{\alpha}\right)$,
where $p$ is the fluid pressure and $\boldsymbol{\delta}$ is the
Kronecker-delta. The boundary integral representation of the Stokes
equation is then used to write the flow produced by an active particle
in an imposed velocity field $\boldsymbol{v}^{\infty}(\boldsymbol{r})$
\citep{lorentz1907,fkg1930bandwertaufgaben,ladyzhenskaya1969,youngren1975stokes,zick1982stokes,muldowney1995spectral,pozrikidis1992,cheng2005heritage,leal2007advanced,singh2015many},
using the Einstein summation convention for repeated Cartesian indices,
\begin{align}
v_{\alpha}(\boldsymbol{r}) & =v_{\alpha}^{\infty}(\boldsymbol{r})-\int G_{\alpha\beta}(\boldsymbol{r},\boldsymbol{r}^{\prime})f_{\beta}(\boldsymbol{r}^{\prime})\,d\mathcal{S}+\int K_{\beta\alpha\nu}(\boldsymbol{r}^{\prime},\boldsymbol{r})\hat{\rho}_{\nu}^{\prime}v_{\beta}(\boldsymbol{r}^{\prime})\,d\mathcal{S},\qquad\boldsymbol{r}\in\mathcal{V},\quad\boldsymbol{r}^{\prime}=\boldsymbol{R}+\boldsymbol{\rho}^{\prime}\in\mathcal{S}.\label{eq:BI}
\end{align}
An outline of the derivation of this classical result is presented
in Appendix \ref{app:Derivation-boundary-integral} in the notation
of this paper. In the above, $\mathcal{V}$ indicates the volume of
the surrounding fluid. The integral kernels for the fluid velocity
are the Green's function $\boldsymbol{G}$ of Stokes flow and the
stress tensor $\boldsymbol{K}$ associated with it. Together with
the pressure field $\boldsymbol{P}$ they satisfy \citep{pozrikidis1992}
\begin{gather}
\nabla_{\alpha}G_{\alpha\beta}(\boldsymbol{r},\boldsymbol{r}^{\prime})=0,\quad-\nabla_{\alpha}P_{\beta}(\boldsymbol{r},\boldsymbol{r}^{\prime})+\eta\nabla^{2}G_{\alpha\beta}(\boldsymbol{r},\boldsymbol{r}^{\prime})=-\delta(\boldsymbol{r}-\boldsymbol{r}^{\prime})\delta_{\alpha\beta},\nonumber \\
K_{\alpha\beta\nu}(\boldsymbol{r},\boldsymbol{r}^{\prime})=-\delta_{\alpha\nu}P_{\beta}(\boldsymbol{r},\boldsymbol{r}^{\prime})+\eta\left[\nabla_{\nu}G_{\alpha\beta}(\boldsymbol{r},\boldsymbol{r}^{\prime})+\nabla_{\alpha}G_{\nu\beta}(\boldsymbol{r},\boldsymbol{r}^{\prime})\right],\label{eq:properties_G}
\end{gather}
where the derivatives are taken with respect to the first argument;
here $\boldsymbol{\nabla}=\boldsymbol{\nabla}_{\boldsymbol{r}}$.
Furthermore, the Green's function satisfies the symmetry $G_{\alpha\beta}(\boldsymbol{r},\boldsymbol{r}^{\prime})=G_{\beta\alpha}(\boldsymbol{r}^{\prime},\boldsymbol{r}).$
By analogy with potential theory, the terms in \eqref{eq:BIE-3parts}
containing the Green's function $\boldsymbol{G}$ and the stress tensor
$\boldsymbol{K}$ are referred to as ``single-layer'' integral and
``double-layer'' integral, respectively \citep{kim2005}. The traction
$\boldsymbol{f}$ is the normal component of the Cauchy stress tensor
evaluated at the surface of the colloid. For $\boldsymbol{r=\boldsymbol{R}}+\boldsymbol{\rho}\in\mathcal{S}$
being evaluated on the surface of the colloid, and thus evaluating
the double-layer integral as a principal value, we have \citep{lorentz1907,fkg1930bandwertaufgaben,ladyzhenskaya1969,youngren1975stokes,zick1982stokes,muldowney1995spectral,pozrikidis1992,cheng2005heritage,leal2007advanced,singh2015many}
\begin{align}
\tfrac{1}{2}v_{\alpha}(\boldsymbol{r}) & =v_{\alpha}^{\infty}(\boldsymbol{r})-\int G_{\alpha\beta}(\boldsymbol{r},\boldsymbol{r}^{\prime})f_{\beta}(\boldsymbol{r}^{\prime})\,d\mathcal{S}+\int K_{\beta\alpha\nu}(\boldsymbol{r}^{\prime},\boldsymbol{r})\hat{\rho}_{\nu}^{\prime}v_{\beta}(\boldsymbol{r}^{\prime})\,d\mathcal{S},\quad\boldsymbol{r},\boldsymbol{r}^{\prime}\in\mathcal{S}.\label{eq:BIE}
\end{align}
This is a Fredholm integral equation of the first kind for the unknown
traction $\boldsymbol{f}$, defined in Eq. \eqref{eq:traction}. By
linearity of Stokes flow, the three distinct contributions to the
traction satisfy \emph{independent} boundary integral equations. These
are
\begin{eqnarray}
v_{\alpha}^{\mathcal{D}}(\boldsymbol{r})=-\int G_{\alpha\beta}(\boldsymbol{r},\boldsymbol{r}^{\prime})f_{\beta}^{\mathcal{D}}(\boldsymbol{r}^{\prime})\,d\mathcal{S}, & \qquad & \text{(rigid body),}\nonumber \\
v_{\alpha}^{\infty}(\boldsymbol{r})=\int G_{\alpha\beta}(\boldsymbol{r},\boldsymbol{r}^{\prime})f_{\beta}^{\infty}(\boldsymbol{r}^{\prime})\,d\mathcal{S}, & \qquad & \text{(imposed flow),}\nonumber \\
\tfrac{1}{2}v_{\alpha}^{\mathcal{A}}(\boldsymbol{r})=-\int\left[G_{\alpha\beta}(\boldsymbol{r},\boldsymbol{r}^{\prime})f_{\beta}^{\mathcal{A}}(\boldsymbol{r}^{\prime})-K_{\beta\alpha\nu}(\boldsymbol{r}^{\prime},\boldsymbol{r})\hat{\rho}_{\nu}^{\prime}v_{\beta}^{\mathcal{A}}(\boldsymbol{r}^{\prime})\right]d\mathcal{S}, & \qquad & \text{(active slip)}.\label{eq:BIE-3parts}
\end{eqnarray}
In writing the rigid body part of Eq. (\ref{eq:BIE-3parts}), we have
used the well-known result that rigid body motion is an eigenfunction
of the double-layer integral operator with eigenvalue $-1/2$ \citep{kim2015ellipsoidal}
{[}see Eqs. \eqref{eq:SL_DL} for a proof{]}. In the following, we
shall solve these integral equations to find the exact solution for
the Stokes traction on an active particle in an arbitrary imposed
flow given in \eqref{eq:scalarStokes}.

\subsection{Exact solution of the boundary integral equation\label{sec:frictionTensors}}

To solve the integral equations \eqref{eq:BIE-3parts} for the unknown
surface tractions, we parametrise the surface fields in terms of TSH
as prescribed in \eqref{eq:expansion-v}. We can use the orthogonality
of the basis functions,
\begin{equation}
\int\mathbf{Y}^{(l)}\mathbf{Y}^{(l')}d\mathcal{S}=\delta_{ll'}\frac{1}{w_{l+1}\tilde{w}_{l+1}}\mathbf{\boldsymbol{\Delta}}^{(l)},\label{eq:orthoTSH}
\end{equation}
to obtain the expansion coefficients
\begin{gather}
\boldsymbol{V}^{\lambda(l)}=\tilde{w}_{l}\int\boldsymbol{v}^{\lambda}(\boldsymbol{R}+\boldsymbol{\rho})\boldsymbol{Y}^{(l-1)}(\hat{\boldsymbol{\rho}})d\mathcal{S},\quad\boldsymbol{F}^{\lambda(l)}=w_{l}\int\boldsymbol{f}^{\lambda}(\boldsymbol{R}+\boldsymbol{\rho})\boldsymbol{Y}^{(l-1)}(\hat{\boldsymbol{\rho}})d\mathcal{S}.\label{eq:coeffModes}
\end{gather}

Having expanded the boundary fields in \eqref{eq:BIE-3parts} in an
orthogonal basis, we use the Ritz-Galerkin method of minimising the
residual to obtain a self-adjoint linear system for the expansion
coefficients \citep{singh2015many,singh2018generalized}. By multiplying
the boundary integral equation by $\boldsymbol{Y}^{(l-1)}(\hat{\boldsymbol{\rho}}),$
and integrating it over the surface of the colloid we obtain the linear
system of equations for the velocity and traction coefficients 
\begin{eqnarray}
\boldsymbol{V}^{\mathcal{D}(l)}=-\boldsymbol{\mathcal{G}}^{(l,l^{\prime})}\odot\mathbf{F}^{\mathcal{D}(l^{\prime})}, & \qquad & \text{(rigid body)},\nonumber \\
\boldsymbol{V}^{\mathcal{\infty}(l)}=\boldsymbol{\mathcal{G}}^{(l,l^{\prime})}\odot\mathbf{F}^{\infty(l^{\prime})}, & \qquad & \text{(imposed flow),}\nonumber \\
\tfrac{1}{2}\boldsymbol{V}^{\mathcal{A}(l)}=-\boldsymbol{\mathcal{G}}^{(l,l^{\prime})}\odot\mathbf{F}^{\mathcal{A}(l^{\prime})}+\boldsymbol{\mathcal{K}}^{(l,l^{\prime})}\odot\boldsymbol{V}^{\mathcal{A}(l^{\prime})}, & \qquad & \text{(active slip)},\label{eq:BIE-3parts-1}
\end{eqnarray}
where the matrix elements $\boldsymbol{\mathcal{G}}^{(l,l^{\prime})}$
and $\boldsymbol{\mathcal{K}}^{(l,l^{\prime})}$ are due to the single-layer
and double-layer, respectively. These matrix elements can be evaluated
exactly for a spherical colloid in an unbounded fluid. The two key
identities necessary for this are the expansion of the reducible symmetric
tensor $\hat{\rho}_{\alpha_{1}}\dots\hat{\rho}_{\alpha_{l}}$ in the
TSH basis \citep{mazur1982,hess2015tensors,singh2018microhydrodynamics}
\begin{equation}
\hat{\rho}_{\alpha_{1}}\dots\hat{\rho}_{\alpha_{l-1}}=\frac{Y_{\alpha_{1}\dots\alpha_{l-1}}^{(l-1)}}{(2l-3)!!}+\frac{1}{2l-3}\sum_{jk\,\text{pairs}}\delta_{\alpha_{j}\alpha_{k}}\frac{Y_{\alpha_{1}\dots\alpha_{j-1}\alpha_{j+1}\dots\alpha_{k-1}\alpha_{k+1}\dots\alpha_{l-1}}^{(l-3)}}{(2l-7)!!}+\mathcal{O}\left(\boldsymbol{Y}^{(l-5)}\right),\label{eq:redY}
\end{equation}
where the big \emph{$\mathcal{O}$} notation stands for terms involving
components of TSH of rank $\leq l-5$, and the expansion of the plane
wave in the TSH basis 
\begin{equation}
e^{i\boldsymbol{k}\cdot\boldsymbol{\rho}}=4\pi b^{2}\sum_{m=1}^{\infty}i^{m-1}w_{m}\tilde{w}_{m}j_{m-1}(k\rho)\,\boldsymbol{Y}^{(m-1)}(\hat{k})\odot\boldsymbol{Y}^{(m-1)}(\hat{\rho}),\label{eq:PWE}
\end{equation}
where $j_{m}(k\rho$) are spherical Bessel functions, $\rho=\left\Vert \boldsymbol{\rho}\right\Vert _{2}=b$,
and $i=\sqrt{-1}$ is the imaginary unit. For the one-body problem,
both the single- and double-layer integrals exhibit singular kernels
and thus the boundary integral equations cannot simply be Taylor expanded
as in \citep{ishikawa2006,swanModelingHydrodynamicSelfpropulsion2011,singh2015many,singh2018generalized}.
However, exploiting translational invariance, we can solve them in
Fourier space. For this, we use the following Fourier representation
of fields $\varphi(\boldsymbol{r})$, 
\begin{equation}
\varphi(\boldsymbol{s})=\int\hat{\varphi}(\boldsymbol{k})e^{i\boldsymbol{k}\cdot\boldsymbol{s}}\frac{d\boldsymbol{k}}{\left(2\pi\right)^{3}},\quad\hat{\varphi}(\boldsymbol{k})=\int\varphi(\boldsymbol{s})e^{-i\boldsymbol{k}\cdot\boldsymbol{s}}d\boldsymbol{s}.
\end{equation}
We now turn to the evaluation of the matrix elements. 

\subsubsection{Single-layer matrix element\label{subsec:Single-layer}}

The single-layer matrix element of Eq. \eqref{eq:BIE-3parts-1} is
given by

\begin{align}
\boldsymbol{\mathcal{G}}^{(l,l^{\prime})} & =\tilde{w}_{l}\tilde{w}_{l'}\int\boldsymbol{Y}^{(l-1)}(\hat{\boldsymbol{\rho}})\boldsymbol{G}(\boldsymbol{r},\boldsymbol{r}^{\prime})\boldsymbol{Y}^{(l^{\prime}-1)}(\hat{\boldsymbol{\rho}}^{\prime})d\mathcal{S}d\mathcal{S}^{\prime},\quad\boldsymbol{r},\boldsymbol{r}^{\prime}\in\mathcal{S}.\label{eq:matrix-element-def-G-1}
\end{align}
In an unbounded fluid we have for the Green's function of Stokes flow
and its Fourier transform \citep{pozrikidis1992}
\begin{equation}
\boldsymbol{G}(\boldsymbol{s})=\frac{1}{8\pi\eta}\frac{1}{s}\left(\boldsymbol{\delta}+\hat{\boldsymbol{s}}\hat{\boldsymbol{s}}\right),\qquad\hat{\boldsymbol{G}}(\boldsymbol{k})=\frac{1}{\eta k^{2}}\left(\boldsymbol{\delta}-\hat{\boldsymbol{k}}\hat{\boldsymbol{k}}\right)=\frac{1}{3\eta k^{2}}\left(2\boldsymbol{\delta}-\boldsymbol{Y}^{(2)}(\hat{\boldsymbol{k}})\right),\label{eq:OseenTensor-1}
\end{equation}
where $\boldsymbol{s}=\boldsymbol{\rho}-\boldsymbol{\rho}^{\prime}$,
and we have used Eq. \eqref{eq:redY}. Using the Fourier transform,
together with the plane wave expansion \eqref{eq:PWE} in the matrix
element \eqref{eq:matrix-element-def-G-1}, we obtain 
\begin{multline}
\mathcal{G}_{\alpha\nu_{1}\dots\nu_{l-1}\beta\kappa_{1}\dots\kappa_{l^{\prime}-1}}^{(l,l^{\prime})}=\sum_{m,m'=1}^{\infty}\tau_{ll'mm'}^{G}\int d\mathcal{S}\,Y_{\nu_{1}\dots\nu_{l-1}}^{(l-1)}(\hat{\rho})Y_{\mu_{1}\dots\mu_{m-1}}^{(m-1)}(\hat{\rho})\int dk\,j_{m-1}(kb)j_{m'-1}(kb)\\
\times\int d\mathcal{S}^{\prime}\,Y_{\kappa_{1}\dots\kappa_{l^{\prime}-1}}^{(l^{\prime}-1)}(\hat{\rho}^{\prime})Y_{\eta_{1}\dots\eta_{m'-1}}^{(m'-1)}(\hat{\rho}^{\prime})\int d\Omega_{k}\,Y_{\mu_{1}\dots\mu_{m-1}}^{(m-1)}(\hat{k})k^{2}\hat{G}_{\alpha\beta}(\boldsymbol{k})Y_{\eta_{1}\dots\eta_{m'-1}}^{(m'-1)}(\hat{k}),
\end{multline}
where $\int d\mathcal{S}$ implies the integral over the surface of
a sphere with radius $b$, $\int d\Omega$ the integral over the surface
of a unit-sphere, and $\int dk$ a scalar definite integral from $0$
to $\infty$, and with
\[
\tau_{ll'mm'}^{G}=\frac{2\eta b^{4}}{\pi}i^{m+3m'}\tilde{w}_{l}\tilde{w}_{l^{\prime}}w_{m}w_{m'}\tilde{w}_{m}\tilde{w}_{m'}.
\]
The integral over the pair of spherical Bessel functions can be found
in \citep{gradshteyn2014table}. With this, the results for surface
integrals over outer products of multiple TSH in \citep{brunnMotionSlightlyDeformed1979},
and the properties of the isotropic tensor $\boldsymbol{\Delta}$
\citep{brunn1976effect,brunnMotionSlightlyDeformed1979,hess2015tensors},
we eventually obtain the result for the single-layer matrix element

\begin{gather}
\mathcal{G}_{\alpha\nu_{1}\dots\nu_{l-1}\beta\kappa_{1}\dots\kappa_{l^{\prime}-1}}^{(l,l^{\prime})}=\delta_{ll^{\prime}}\mathcal{G}_{0}^{(l)}\left[\delta_{\alpha\beta}\Delta_{\nu_{1}\dots\nu_{l-1},\kappa_{1}\dots\kappa_{l-1}}^{(l-1)}-\frac{l\left(2l-1\right)}{2\left(l-1\right)\left(2l+1\right)}\,\,\Lambda_{\alpha\nu_{1}\dots\nu_{l-1}\beta\kappa_{1}\dots\kappa_{l-1}}^{(l)}\right].\label{eq:SL-1}
\end{gather}
Here, $\mathcal{G}_{0}^{(l)}=(l-1)^{2}/(2\pi\eta bw_{l-1})$ and $\Lambda_{\alpha\nu_{1}\dots\nu_{l-1}\beta\kappa_{1}\dots\kappa_{l-1}}^{(l)}=$
$\Delta_{\nu_{1}\dots\nu_{l-1}\alpha,\beta\kappa_{1}\dots\kappa_{l-1}}^{(l)}+$
$\Delta_{\nu_{1}\dots\nu_{l-1}\beta,\alpha\kappa_{1}\dots\kappa_{l-1}}^{(l)}$.

\subsubsection{Double-layer matrix element\label{subsec:Double-layer}}

The double-layer matrix element of Eq. \eqref{eq:BIE-3parts-1} is
given by

\begin{align}
\boldsymbol{\mathcal{K}}^{(l,l^{\prime})} & =\tilde{w}_{l}w_{l'}\int\boldsymbol{Y}^{(l-1)}(\hat{\boldsymbol{\rho}})\boldsymbol{K}(\boldsymbol{r}^{\prime},\boldsymbol{r})\cdot\hat{\boldsymbol{\rho}}^{\prime}\boldsymbol{Y}^{(l^{\prime}-1)}(\hat{\boldsymbol{\rho}}^{\prime})d\mathcal{S}d\mathcal{S}^{\prime},\quad\boldsymbol{r},\boldsymbol{r}^{\prime}\in\mathcal{S}.\label{eq:matrix-element-def-K-1}
\end{align}
In the unbounded domain the stress tensor corresponding to the Green's
function \eqref{eq:OseenTensor-1} and its Fourier transform are \citep{pozrikidis1992}
\begin{equation}
\boldsymbol{K}(\boldsymbol{s})=-\frac{3}{4\pi}\frac{1}{s^{2}}\hat{\boldsymbol{s}}\hat{\boldsymbol{s}}\hat{\boldsymbol{s}},\qquad\hat{\boldsymbol{K}}(\boldsymbol{k})=\frac{2i}{k}\left[3\left(\hat{\boldsymbol{k}}\boldsymbol{\delta}\right)^{\text{sym}}-\hat{\boldsymbol{k}}\hat{\boldsymbol{k}}\hat{\boldsymbol{k}}\right]=\frac{2i}{5k}\left[9\left(\boldsymbol{Y}^{(1)}(\hat{\boldsymbol{k}})\boldsymbol{\delta}\right)^{\text{sym}}-\frac{1}{3}\boldsymbol{Y}^{(3)}(\hat{\boldsymbol{k}})\right],\label{eq:StressTensor}
\end{equation}
where $\boldsymbol{s}=\boldsymbol{\rho}-\boldsymbol{\rho}^{\prime}$,
and the notation $\left(\dots\right)^{\text{sym}}$ implies a projection
onto the symmetric part of the tensor, eg $\left(\hat{\rho}_{\alpha}\hat{\rho}'_{\beta}\right)^{\text{sym}}$$=\tfrac{1}{2}\left(\hat{\rho}_{\alpha}\hat{\rho}'_{\beta}+\hat{\rho}_{\beta}\hat{\rho}'_{\alpha}\right)$.
We have once again used Eq. \eqref{eq:redY}. Using this Fourier transform
and the plane wave expansion \eqref{eq:PWE} in the matrix element
\eqref{eq:matrix-element-def-K-1} the expression for the double-layer
matrix element becomes
\begin{multline}
K_{\alpha\nu_{1}\dots\nu_{l-1}\beta\kappa_{1}\dots\kappa_{l^{\prime}-1}}^{(l,l^{\prime})}=\sum_{m,m'=1}^{\infty}\tau_{ll'mm'}^{K}\int d\mathcal{S}\,Y_{\nu_{1}\dots\nu_{l-1}}^{(l-1)}(\hat{\rho})Y_{\mu_{1}\dots\mu_{m-1}}^{(m-1)}(\hat{\rho})\int dk\,kj_{m-1}(kb)j_{m'-1}(kb)\\
\times\int d\mathcal{S}^{\prime}\,\hat{\rho}_{\eta}^{\prime}Y_{\kappa_{1}\dots\kappa_{l^{\prime}-1}}^{(l^{\prime}-1)}(\hat{\rho}^{\prime})Y_{\eta_{1}\dots\eta_{m'-1}}^{(m'-1)}(\hat{\rho}^{\prime})\int d\Omega_{k}\,Y_{\mu_{1}\dots\mu_{m-1}}^{(m-1)}(\hat{k})k\hat{K}_{\beta\alpha\eta}(\boldsymbol{k})Y_{\eta_{1}\dots\eta_{m'-1}}^{(m'-1)}(\hat{k})
\end{multline}
with 
\[
\tau_{ll'mm'}^{K}=\frac{2b^{4}}{\pi}i^{m'+3m}\tilde{w}_{l}w_{l^{\prime}}w_{m}w_{m'}\tilde{w}_{m}\tilde{w}_{m'}.
\]
Again, the relevant integral over spherical Bessel functions can be
found in \citep{gradshteyn2014table}. Using the results for integrals
over multiple TSH obtained by Brunn \citep{brunnMotionSlightlyDeformed1979},
and the properties of the $\boldsymbol{\Delta}$-tensor, we find the
double-layer matrix element after lengthy manipulation, 

\begin{gather}
\mathcal{K}_{\alpha\nu_{1}\dots\nu_{l-1}\beta\kappa_{1}\dots\kappa_{l^{\prime}-1}}^{(l,l^{\prime})}=\delta_{ll^{\prime}}\mathcal{K}_{0}^{(l)}\left[\delta_{\alpha\beta}\Delta_{\nu_{1}\dots\nu_{l-1},\kappa_{1}\dots\kappa_{l-1}}^{(l-1)}-\frac{2l}{2l+1}\,\Lambda_{\alpha\nu_{1}\dots\nu_{l-1}\beta\kappa_{1}\dots\kappa_{l-1}}^{(l)}\right],\label{eq:DL-1}
\end{gather}
where $\mathcal{K}_{0}^{(l)}=3/\left(4l-6\right)$. 

\subsection{Diagonalisation of the linear system of equations\label{sec:Stokes-traction}}

In the following, we explicitly define the irreducible representation
of the coefficients of the boundary velocity and traction in \eqref{eq:decomp-proj}.
We then use this to project the linear system \eqref{eq:BIE-3parts-1}
onto its irreducible subspaces. By doing so, the linear system diagonalises
and thus can be solved trivially. This results directly in the generalised
Stokes laws in Eq. \eqref{eq:scalarStokes}.

As is evident from the single-layer, Eq. \eqref{eq:SL-1}, and double-layer,
Eq. \eqref{eq:DL-1}, matrix elements, the linear system \eqref{eq:BIE-3parts-2}
naturally diagonalises in the modes $(l)$ of the expansion coefficients.
We will now show that it is in fact diagonal in all its irreducible
subspaces. First, we define the decomposition operators used in Eq.
\eqref{eq:decomp-proj} as 
\begin{align}
 & \left[\boldsymbol{D}^{(ls)}\odot\boldsymbol{F}^{\lambda(ls)}\right]_{\alpha\nu_{1}\dots\nu_{l-1}}\,=\Delta_{\alpha\nu_{1}\dots\nu_{l-1},\beta\kappa_{1}\dots\kappa_{l-1},}^{(l)}F_{\beta\kappa_{1}\dots\kappa_{l-1}}^{\lambda(ls)},\nonumber \\
 & \left[\boldsymbol{D}^{(la)}\odot\boldsymbol{F}^{\lambda(la)}\right]_{\alpha\nu_{1}\dots\nu_{l-1}}=-\tfrac{l-1}{l}\Delta_{\nu_{1}\dots\nu_{l-1},\kappa_{1}\dots\kappa_{l-1}}^{(l-1)}\epsilon_{\alpha\kappa_{1}\beta}F_{\beta\kappa_{2}\dots\kappa_{l-1}}^{\lambda(la)},\nonumber \\
 & \left[\boldsymbol{D}^{(lt)}\odot\boldsymbol{F}^{\lambda(lt)}\right]_{\alpha\nu_{1}\dots\nu_{l-1}}\,\,=\tfrac{2l-3}{2l-1}\Delta_{\nu_{1}\dots\nu_{l-1},\kappa_{1}\dots\kappa_{l-1}}^{(l-1)}\delta_{\alpha\kappa_{1}}F_{\kappa_{2}\dots\kappa_{l-1}}^{\lambda(lt)},\label{eq:projection_operator-1-1}
\end{align}
with the corresponding projection operators $\boldsymbol{P}^{(l\sigma)}$
\begin{align}
 & \left[\boldsymbol{P}^{(ls)}\odot\boldsymbol{F}^{\lambda(l)}\right]_{\beta\kappa_{1}\cdots\kappa_{l-1}}=\,\Delta_{\beta\kappa_{1}\cdots\kappa_{l-1},\sigma\mu_{1}\dots\mu_{l-1}}^{(l)}F_{\sigma\mu_{1}\dots\mu_{l-1}}^{\lambda(l)},\quad\,\,\nonumber \\
 & \left[\boldsymbol{P}^{(la)}\odot\boldsymbol{F}^{\lambda(l)}\right]_{\lambda\kappa_{2}\dots\kappa_{l-1}}=\,\Delta_{\lambda\kappa_{2}\dots\kappa_{l-1},\mu\eta_{2}\dots\eta_{l-1}}^{(l-1)}\epsilon_{\mu\beta\alpha}F_{\alpha\beta\eta_{2}\dots\eta_{l-1}}^{\lambda(l)},\quad\,\,\nonumber \\
 & \left[\boldsymbol{P}^{(lt)}\odot\boldsymbol{F}^{\lambda(l)}\right]_{\kappa_{2}\dots\kappa_{l-1}}\,\,\,\,=\,\delta_{\mu\lambda}F_{\mu\lambda\kappa_{2}\dots\kappa_{l-1}}^{\lambda(l)}.\quad\,\,\label{eq:projection_operator-2}
\end{align}
Here, $\boldsymbol{\epsilon}$ is the Levi-Civita tensor and $\boldsymbol{\delta}$
is the Kronecker delta. With this we can define what we call the
``irreducible matrix elements'' 
\begin{equation}
\boldsymbol{\mathcal{G}}^{(l\sigma,l^{\prime}\sigma^{\prime})}=\boldsymbol{P}^{(l\sigma)}\odot\boldsymbol{\mathcal{G}}^{(l,l^{\prime})}\odot\boldsymbol{D}^{(l'\sigma')},\qquad\boldsymbol{\mathcal{K}}^{(l\sigma,l^{\prime}\sigma^{\prime})}=\boldsymbol{P}^{(l\sigma)}\odot\boldsymbol{\mathcal{K}}^{(l,l^{\prime})}\odot\boldsymbol{D}^{(l'\sigma')}.\label{eq:irredMatElms}
\end{equation}
Using these in the linear system \eqref{eq:BIE-3parts-1}, the result
is a self-adjoint linear system in the irreducible expansion coefficients,
\begin{eqnarray}
\boldsymbol{V}^{\mathcal{D}(l\sigma)}=-\boldsymbol{\mathcal{G}}^{(l\sigma,l^{\prime}\sigma^{\prime})}\odot\mathbf{F}^{\mathcal{D}(l^{\prime}\sigma^{\prime})}, & \qquad & \text{(rigid body)}\nonumber \\
\boldsymbol{V}^{\infty(l\sigma)}=\boldsymbol{\mathcal{G}}^{(l\sigma,l^{\prime}\sigma^{\prime})}\odot\mathbf{F}^{\infty(l^{\prime}\sigma^{\prime})}, & \qquad & \text{(imposed flow)}\nonumber \\
\tfrac{1}{2}\boldsymbol{V}^{\mathcal{A}(l\sigma)}=-\boldsymbol{\mathcal{G}}^{(l\sigma,l^{\prime}\sigma^{\prime})}\odot\mathbf{F}^{\mathcal{A}(l^{\prime}\sigma^{\prime})}+\boldsymbol{\mathcal{K}}^{(l\sigma,l^{\prime}\sigma^{\prime})}\odot\boldsymbol{V}^{\mathcal{A}(l^{\prime}\sigma^{\prime})}. & \qquad & \text{(active slip)}\label{eq:BIE-3parts-2}
\end{eqnarray}
Inserting the matrix elements \eqref{eq:SL-1} and \eqref{eq:DL-1},
together with the definitions of the decomposition \eqref{eq:projection_operator-1-1}
and projection \eqref{eq:projection_operator-2} operators, into Eq.
\eqref{eq:irredMatElms}, it is straightforward to show that
\begin{gather}
\boldsymbol{\mathcal{G}}^{(l\sigma,l'\sigma')}\odot\boldsymbol{F}^{\mathcal{\lambda}(l'\sigma')}=\delta_{ll'}\delta_{\sigma\sigma'}\,\,g_{l\sigma}\boldsymbol{F}^{\lambda(l\sigma)},\qquad\boldsymbol{\mathcal{K}}^{(l\sigma,l'\sigma')}\odot\boldsymbol{V}^{\mathcal{\lambda}(l'\sigma')}=\delta_{ll'}\delta_{\sigma\sigma'}\,\,k_{l\sigma}\boldsymbol{V}^{\lambda(l\sigma)}\label{eq:SL_DL}
\end{gather}
Here, the scalar $l$-dependent coefficients $g_{l\sigma}$ and $k_{l\sigma}$
are
\begin{align}
 & g_{ls}=\frac{l+1}{2l+1}g_{la}, &  & g_{la}=\frac{1}{4\pi\eta b\,w_{l}}, &  & g_{lt}=\frac{l-2}{2l-3}g_{la},\nonumber \\
 & k_{ls}=\frac{1}{2l+1}k_{la}, &  & k_{la}=-\frac{3}{2(2l-1)}, &  & k_{lt}=-\frac{1}{2l-3}k_{la}.\label{eq:SLDLcoeffs}
\end{align}
It is worth noting that both single- and double-layer irreducible
matrix elements vanish identically for non-diagonal combinations of
modes $(l\sigma,l^{\prime}\sigma^{\prime})$, i.e., when $l\sigma\neq l^{\prime}\sigma^{\prime}$,
apart from $(l\sigma,l^{\prime}\sigma^{\prime})=(lt,la)$. However,
upon contraction with an irreducible tensor these too vanish, i.e.,
$\boldsymbol{\mathcal{G}}^{(lt,la)}\odot\boldsymbol{F}^{(la)}=0$
and $\boldsymbol{\mathcal{K}}^{(lt,la)}\odot\boldsymbol{V}^{(la)}=0.$
Thus the linear system arising from \eqref{eq:BIE-3parts} is diagonal
not only in $(l)$, but also in all its irreducible subspaces labelled
by $(l\sigma)$.

Using this diagonal solution for the linear system \eqref{eq:BIE-3parts-2},
we can straightforwardly write down the generalised Stokes laws in
\eqref{eq:scalarStokes} for an isolated active particle in an unbounded
domain. To summarise, we have derived exact expressions for the friction
coefficients $\gamma_{l\sigma}$ and $\hat{\gamma}_{l\sigma}$ due
to imposed flow and active surface slip, obtained using the direct
formulation of the boundary integral equation.

\section{Applications\label{sec:stress}}

In this section we briefly discuss some applications of the above
results. First, the connection of our results with the generalised
Fax{\'e}n  relations are made explicit. Subsequently, we express
the irreducible expansion coefficients $\boldsymbol{V}^{\lambda(l\sigma)}$
and $\boldsymbol{F}^{\lambda(l\sigma)}$ in terms of standard physical
quantities. In doing so, we make the observation that the symmetric-irreducible
dipole on the particle depends on whether it is strained by its active
surface slip or by an imposed shear flow. We then obtain a simple
expression for the power dissipation of an active colloid in an imposed
flow in terms of the generalised friction coefficients in \eqref{eq:scalarFriction}
and the modes $\boldsymbol{V}^{\lambda(l\sigma)}$. Finally, we consider
thermal fluctuations in the fluid and the associated traction modes
acting on the particle. Using the diagonalisation of the matrix elements
in a basis of TSH \eqref{eq:SL_DL} and the results in \citep{singh2017fluctuation}
on fluctuating hydrodynamics, we give an explicit expression for the
variance of the fluctuation traction modes. 

\subsection{Generalised Fax{\'e}n  relations\label{subsec:Expansion-coefficients-vs}}

Here, we derive the relation between a Taylor expansion of the imposed
flow about the centre of the particle and its expansion coefficients.
The derivation for a similar relation regarding the boundary integral
of the Green's function can be found in \citep{singh2015many}. The
expansion coefficients $\boldsymbol{V}^{\infty(l)}$ are defined in
\eqref{eq:coeffModes}. The Taylor expansion of the imposed flow about
the centre of the sphere is given as 
\[
\boldsymbol{v}^{\infty}(\boldsymbol{R}+\boldsymbol{\rho})=\sum_{l=1}^{\infty}\frac{1}{(l-1)!}\left(\boldsymbol{\rho}\cdot\boldsymbol{\nabla}\right)^{(l-1)}\left.\boldsymbol{v}^{\infty}(\boldsymbol{R}+\boldsymbol{\rho})\right|_{\boldsymbol{\rho}=\boldsymbol{0}},
\]
where we have defined
\[
\left(\boldsymbol{\rho}\cdot\boldsymbol{\nabla}\right)^{(l-1)}=\rho_{\alpha_{1}}\rho_{\alpha_{2}}\dots\rho_{\alpha_{l-1}}\nabla_{\alpha_{1}}\nabla_{\alpha_{2}}\dots\nabla_{\alpha_{l-1}}.
\]
Using \eqref{eq:redY} we can write this in terms of TSH
\[
\left(\boldsymbol{\rho}\cdot\boldsymbol{\nabla}\right)^{(l-1)}=b^{l-1}\left[\frac{\boldsymbol{Y}^{(l-1)}\odot\boldsymbol{\nabla}^{(l-1)}}{(2l-3)!!}+\frac{1}{2l-3}\sum_{jk\,\text{pairs}}\frac{\boldsymbol{Y}^{(l-3)}\odot\boldsymbol{\nabla}^{(l-3)}\nabla^{2}}{(2l-7)!!}+\mathcal{O}\left(\boldsymbol{Y}^{(l-5)}\odot\boldsymbol{\nabla}^{(l-5)}\right)\right].
\]
Due to orthogonality of the TSH, only two terms remain upon integration
over the surface of the sphere in the definition of the expansion
coefficients \eqref{eq:coeffModes}. In the irreducible subspaces,
therefore, a Taylor expansion of the imposed flow and its expansion
coefficients are related as
\begin{equation}
\boldsymbol{V}^{\infty(l\sigma)}=\boldsymbol{P}^{(l\sigma)}\odot\left[b^{l-1}\boldsymbol{\Delta}^{(l-1)}\left(1+\frac{b^{2}}{4l+2}\nabla^{2}\right)\boldsymbol{\nabla}^{(l-1)}\boldsymbol{v}^{\infty}\right]_{\boldsymbol{R}}.\label{eq:taylor_yl}
\end{equation}
Here $[\dots]_{\boldsymbol{R}}$ denotes that the function inside
the bracket is evaluated at the centre $\boldsymbol{R}$ of the particle.
In this paper, we have used the approach to expand the boundary fields
in TSH for both imposed flow and active surface slip. It should be
noted that a corresponding Taylor expansion about the centre of the
particle is not possible for the active slip which is only defined
at the surface of the particle. Using a different method, Brunn \citep{brunn1980faxen}
has obtained relations analogous to \eqref{eq:taylor_yl} and termed
them Fax{\'e}n  relations.

\subsection{Symmetric-irreducible dipole and stresslet\label{subsec:SymIrrDipole}}

With \eqref{eq:coeffModes} and \eqref{eq:decomp-proj} the irreducible
expansion coefficients are readily expressed in terms of commonly
used physical quantities. The rigid body motion velocity expansion
coefficients $\boldsymbol{V}^{\mathcal{D}(l\sigma)}$ only have two
non-vanishing modes, corresponding to translational velocity $\boldsymbol{V}=\boldsymbol{V}^{\mathcal{D}(1s)}$
and rotational velocity $\boldsymbol{\Omega}=\boldsymbol{V}^{\mathcal{D}(2a)}/2b$,
with $\boldsymbol{V}^{\mathcal{D}(l\sigma)}=0\,\forall\,l\sigma\notin\left\{ 1s,2a\right\} $.
Similarly, the first two modes of the imposed flow are $\boldsymbol{V}^{\infty(1s)}=\boldsymbol{V}^{\infty}$
and $\boldsymbol{V}^{\infty(2a)}=2b\boldsymbol{\Omega}^{\infty}$,
while the first two modes of activity are $\boldsymbol{V}^{\mathcal{A}(1s)}=-\boldsymbol{V}^{\mathcal{A}}$
and $\boldsymbol{V}^{\mathcal{A}(2a)}=-2b\boldsymbol{\Omega}^{\mathcal{A}}$.
Here, the active translational velocity $\boldsymbol{V}^{\mathcal{A}}$
and the active angular velocity $\boldsymbol{\Omega}^{\mathcal{A}}$
of a spherical active particle \citep{anderson1991,stone1996,ghose2014irreducible}
are given by
\begin{gather}
\boldsymbol{V}^{\mathcal{A}}=-\frac{1}{4\pi b^{2}}\int\boldsymbol{v}^{\mathcal{A}}(\boldsymbol{\rho})d\mathcal{S},\qquad\boldsymbol{\Omega}^{\mathcal{A}}=-\frac{3}{8\pi b^{4}}\int\boldsymbol{\rho}\times\boldsymbol{v}^{\mathcal{A}}(\boldsymbol{\rho})d\mathcal{S}.\label{eq:avgVel}
\end{gather}
We also define the rate of strain dyadic $\boldsymbol{E}^{\lambda}=\boldsymbol{V}^{\lambda(2s)}/b$,
due to activity or imposed flow, as
\begin{gather}
E_{\alpha\beta}^{\lambda}=\frac{3}{8\pi b}\int\left(\hat{\rho}_{\alpha}v_{\beta}^{\lambda}+v_{\alpha}^{\lambda}\hat{\rho}_{\beta}\right)d\mathcal{S}.\label{eq:strain-rate}
\end{gather}
Analogously, we identify the most commonly used traction tensors produced
by the corresponding velocity fields
\begin{gather}
\boldsymbol{F}^{\lambda(1s)}=\boldsymbol{F}^{\lambda},\qquad\boldsymbol{F}^{\lambda(2a)}=\frac{1}{b}\boldsymbol{T}^{\lambda},\qquad\boldsymbol{F}^{\lambda(2s)}=\frac{1}{b}\boldsymbol{S}^{\lambda},\label{eq:physicalTrac}
\end{gather}
where $\boldsymbol{F},$ $\boldsymbol{T}$ and $\boldsymbol{S}$ are
the familiar hydrodynamic force and torque and the symmetric-irreducible
second moment of the traction, the symmetric-irreducible dipole. The
latter is 
\begin{equation}
S_{\alpha\beta}^{\lambda}=\int\left[\tfrac{1}{2}\left(f_{\alpha}^{\lambda}\rho_{\beta}+f_{\beta}^{\lambda}\rho_{\alpha}\right)-\tfrac{\delta_{\alpha\beta}}{3}f_{\nu}^{\lambda}\rho_{\nu}\right]d\mathcal{S}.\label{eq:stresslet}
\end{equation}
We note that this is different from the combination of traction and
velocity $(2s)$ mode, first introduced by Landau and Lifshitz \citep{landau1959fluid},
and subsequently called the stresslet by Batchelor \citep{batchelor1970stress}
and derived by various authors since, where more recent derivations
include \citep{ishikawa2006,swanModelingHydrodynamicSelfpropulsion2011,laugaStressletsInducedActive2016,nasouriHigherorderForceMoments2018}.
While Batchelor's stresslet is defined as the contribution of a particle
to the bulk stress, the above, $\boldsymbol{S}^{\lambda}$, describes
the hydrodynamic stress experienced by the particle itself, either
due to its active surface slip or due to an imposed shear flow. In
particular, we want to draw the attention of the reader to the differing
symmetric-irreducible dipoles acting on the colloid, depending on
whether an imposed straining flow $\boldsymbol{E}^{\infty}$ or a
straining flow due to the active surface slip $\boldsymbol{E}^{\mathcal{A}}$
is applied:
\begin{gather}
\boldsymbol{S}^{\infty}=\frac{20\pi\eta b^{3}}{3}\boldsymbol{E}^{\infty},\qquad\boldsymbol{S}^{\mathcal{A}}=-4\pi\eta b^{3}\boldsymbol{E}^{\mathcal{A}}.\label{eq:stresses}
\end{gather}
This result is readily explained by the contribution of the double-layer
integral in \eqref{eq:BIE-3parts} for the active surface slip velocity.
Using the above correspondences between the irreducible modes and
the velocity (angular velocity) and the force (torque) in the generalised
Stokes laws \eqref{eq:scalarFriction} we correctly recover Stokes
law for the translation (rotation) of a spherical object in a viscous
fluid with a friction coefficient of $6\pi\eta b$ ($8\pi\eta b^{3})$. 

\subsection{Power dissipation\label{subsec:Power-dissipation}}

The power dissipation in the volume of the fluid is $\dot{\mathcal{E}}=\int\boldsymbol{\sigma}\colon\left(\boldsymbol{\nabla}\boldsymbol{v}\right)\,d\mathcal{V}$
\citep{landau1959fluid}. Using the divergence theorem to rewrite
this as an integral over the surface of the sphere, we obtain for
the power dissipation due to an active colloid in an imposed flow
\begin{multline}
\dot{\mathcal{E}}=\sum_{\lambda,\lambda'}\left(\gamma_{ls}^{\lambda}\,\,\boldsymbol{V}^{\lambda(ls)}\odot\boldsymbol{V}^{\lambda'(ls)}+\tfrac{2}{2l-3}\left(\tfrac{l-1}{l}\right)^{2}\gamma_{la}^{\lambda}\,\,\boldsymbol{V}^{\lambda(la)}\odot\boldsymbol{V}^{\lambda'(la)}+\tfrac{2l-3}{2l-1}\,\gamma_{lt}^{\lambda}\,\,\boldsymbol{V}^{\lambda(lt)}\odot\boldsymbol{V}^{\lambda'(lt)}\right),\\
\text{where }\gamma_{l\sigma}^{\lambda}=\begin{cases}
\gamma_{l\sigma}, & \text{for }\lambda\in\left\{ \mathcal{D},\infty\right\} ,\\
\hat{\gamma}_{l\sigma}, & \text{for }\lambda=\mathcal{A}.
\end{cases}\label{eq:powerDiss}
\end{multline}
Here, we sum over both, $\lambda,\,\lambda'\in\left\{ \mathcal{D},\infty,\mathcal{A}\right\} $,
while the sum over the existing modes $(l\sigma)$ is left implicit.
In obtaining this result, we have used the generalised Stokes laws
\eqref{eq:scalarStokes}. It correctly follows that $\dot{\mathcal{E}}\geq0$,
i.e., the power dissipation is always positive definite. It is readily
checked that we recover the correct result for the power dissipation
due to rigid body motion, $\dot{\mathcal{E}}^{\mathcal{D}}=6\pi\eta b\,\boldsymbol{V}\cdot\boldsymbol{V}+8\pi\eta b^{3}\,\boldsymbol{\Omega}\cdot\boldsymbol{\Omega}$.
In Appendix \ref{app:Power-dissipation} we simplify the result for
the power dissipation due to active slip only, $\dot{\mathcal{E}}^{\mathcal{A}}$,
further by making use of the uniaxial parametrisation introduced in
the caption of Figure \ref{fig:Slip-and-traction}. 

\subsection{Fluctuating hydrodynamics}

So far, we have ignored the role of thermal fluctuations in the fluid.
At a non-zero temperature $k_{B}T$, considering thermal fluctuations
of the surrounding fluid, we must rewrite Eq. \eqref{eq:traction}
as
\begin{equation}
\boldsymbol{f}=\boldsymbol{f}^{\mathcal{D}}+\boldsymbol{f}^{\infty}+\boldsymbol{f}^{\mathcal{A}}+\boldsymbol{f}^{\mathcal{B}},\label{eq:Brownian-traction}
\end{equation}
where the term $\boldsymbol{f}^{\mathcal{B}}$ now captures the fluctuating
contribution to the traction \citep{hauge1973fluctuating,fox1970contributions,bedeaux1974brownian,roux1992brownian,zwanzig1964hydrodynamic}.
By linearity of Stokes flow, this contribution can be solved for independently.
The Brownian traction is a zero-mean Gaussian random variable and
so it is of particular interest to find an explicit expression for
its variance. Using the fluctuation-dissipation relation, and an expansion
of $\boldsymbol{f}^{\mathcal{B}}$ in TSH analogous to \eqref{eq:expansion-v},
\citep{singh2017fluctuation} have found a formal expression for the
variance of the irreducible fluctuating traction modes $\boldsymbol{F}^{\mathcal{B}(l\sigma)}$
by ``projecting out'' the fluid using the boundary-domain integral
representation of Stokes flow. We can now use the results of Section
\ref{sec:Stokes-traction} to write the variance of these zero-mean
Gaussian random modes explicitly for an active colloid in an unbounded
thermally fluctuating system. The explicit form of the variance for
the fluctuating traction is then
\begin{equation}
\left\langle \boldsymbol{F}^{\mathcal{B}(l\sigma)}(t)\boldsymbol{F}^{\mathcal{B}(l'\sigma')}(t')\right\rangle =\delta_{ll'}\delta_{\sigma\sigma'}\,2k_{B}T\,\delta(t-t')\begin{cases}
\gamma_{ls}\,\boldsymbol{\Delta}^{(l)}, & \text{for }\sigma=s,\\
\tfrac{2l-3}{2}\left(\tfrac{l}{l-1}\right)^{2}\gamma_{la}\,\text{\ensuremath{\boldsymbol{\Delta}^{(l-1)}}}, & \text{for }\sigma=a,\\
\tfrac{2l-1}{2l-3}\,\gamma_{lt}\,\boldsymbol{\Delta}^{(l-2)}, & \text{for }\sigma=t,
\end{cases}\label{eq:variance}
\end{equation}
where $\gamma_{l\sigma}$ are the scalar friction coefficients in
\eqref{eq:scalarFriction}. With this we readily recover the well-known
variances for the Brownian force and torque \citep{zwanzig1964hydrodynamic,chow_simultaneous_1973}.
Canonically denoting the moments of the fluctuating traction by the
superscript $\lambda=\mathcal{B}$, we obtain
\begin{align*}
\left\langle F_{\alpha}^{\mathcal{B}}(t)F_{\beta}^{\mathcal{B}}(t')\right\rangle  & =2k_{B}T\,6\pi\eta b\,\delta_{\alpha\beta}\delta(t-t'),\\
\\
\left\langle T_{\alpha}^{\mathcal{B}}(t)T_{\beta}^{\mathcal{B}}(t')\right\rangle  & =2k_{B}T\,8\pi\eta b^{3}\,\delta_{\alpha\beta}\delta(t-t').
\end{align*}
Furthermore, we give the variance of the fluctuating symmetric-irreducible
dipole
\[
\left\langle S_{\alpha\beta}^{\mathcal{B}}(t)S_{\gamma\kappa}^{\mathcal{B}}(t')\right\rangle =2k_{B}T\,\,\frac{10\pi\eta b^{3}}{3}\,\left(\delta_{\alpha\gamma}\delta_{\beta\kappa}+\delta_{\alpha\kappa}\delta_{\beta\gamma}-\tfrac{2}{3}\delta_{\alpha\beta}\delta_{\gamma\kappa}\right)\delta(t-t').
\]

\section{Conclusion and outlook\label{sec:Discussion}}

We used the direct boundary integral formulation of the Stokes equation
and Ritz-Galerkin discretisation in a basis of tensorial spherical
harmonics to simultaneously diagonalise the single-layer and double-layer
integral operators and, thereby, obtain an exact solution for the
traction on a spherical active particle in an unbounded fluid. The
central result of this paper, Eq. (\ref{eq:scalarStokes}), are expressions
for the linear response of an arbitrary traction mode to a forcing
by the corresponding mode of the active slip and the imposed flow.
We call these linear relations generalised Stokes laws. 

The boundary integral formulation of Stokes flow, spectral expansion
of the surface fields in a basis of polynomials, scalar or vector
spherical harmonics, and Ritz-Galerkin discretisation are classical
methods in computing the slow viscous flows of colloidal particles
\citep{youngren1975stokes,felderhof1976,felderhofForceDensityInduced1976a,zick1982stokes,schmitzCreepingFlowSpherical1982,cichocki1994friction,cichocki_friction_2000,corona2017integralequation}.
It is then worthwhile to compare our main results in Sections \ref{sec:frictionTensors}
and \ref{sec:Stokes-traction} to related work in the literature.
In Table \ref{tab:history} we have listed some important contributions
in chronological order that have (a) analytically obtained the traction
on a single spherical particle in unbounded Stokes flow or (b), alternatively,
obtained the flow field around such a particle, from which the stress
tensor and thus the traction can be derived. This list by no means
is exhaustive and is meant as a chronological overview, rather than
a collection of every relevant contribution to the field. The present
paper fits into the context of other related previous work by some
of the authors \citep{singh2015many,singh2018generalized} as follows.
Despite the treatment of rather general problems, such as many-body
problems in arbitrary confining geometries, the simplest possible
system of a single active colloid in an unbounded and arbitrary imposed
flow was not solved. In particular, it was not known whether the single-
and double-layer integral operators could be diagonalised simultaneously
in a basis of TSH. The present work completes these developments that
follow from \citep{singh2015many,singh2018generalized}.

In future work, we will extend our calculations to obtain explicit
results for the traction on an active particle near surfaces such
as an infinite plane no-slip wall or fluid-fluid interface \citep{blake1971c,felderhof1976,felderhofForceDensityInduced1976a,aderogba1978action,lee_motion_1979,yang_particle_1984,falade_hydrodynamic_1986,cichocki_friction_2000,swan2007simulation,blawzdziewicz_motion_2010,liu_wall_2010,singh2017fluctuation}.
The exact one-body solution presented here will be particularly useful
in obtaining efficient iterative numerical solutions of the boundary
integral equation for many particles \citep{zick1982stokes,ladd1988,brady1988stokesian,ichiki2002improvement,swan2007simulation,fioreRapidSamplingStochastic2018,singh2015many,ishikawa2006,swanModelingHydrodynamicSelfpropulsion2011}.
In this case, the one-body solution can be used to initialise iterations
that converge to the diagonally dominant numerical solutions \citep{singh2018generalized}.
The complete set of modes of the traction derived here can also be
used to study the rheology of active suspensions \citep{batchelor1970stress,brunn1976effect,ishikawa2007rheology}.
In a many-body setting, using TSH as a basis for expansion of the
surface fields has the additional advantage of the basis functions
and the expansion coefficients being irreducible with respect to rotations
\citep{damour_multipole_1991,applequist_maxwellcartesian_2002}. This
allows for the simplified use of the rotation based fast multipole
method (FMM) in summing long-ranged harmonics \citep{greengard1987fast,white_rotating_1996,dachsel_fast_2006,shanker2007accelerated}.
So far, our approach is limited to spherical particles. There exist
a number of papers that have extended related analyses to close-to-spherical
\citep{brunnMotionSlightlyDeformed1979}, ellipsoidal \citep{laugaStressletsInducedActive2016},
and arbitrarily shaped particles \citep{youngren1975stokes,power1987second,stone1996,nasouriHigherorderForceMoments2018},
the latter of which tend to use either the reciprocal theorem to obtain
quite general results, or numerics. In future work, we will aim to
extend our results of arbitrary order to more complex particle shapes.
All of these directions present exciting avenues for future work on
the mechanics and statistical mechanics of active colloidal suspensions.
\begin{acknowledgments}
We thank an anonymous referee for suggestions to improve the presentation
of our results. This work was funded in part by the Engineering and
Physical Sciences Research Council (G.T., project Reference No. 2089780),
the European Research Council under the EU's Horizon 2020 Program
(R.S., ERC Grant Agreement No. 740269), and by an Early Career Grant
to R.A. from the Isaac Newton Trust. The Scientific colour map bamako
\citep{crameri_scientific_2021} is used in this study to prevent
visual distortion of the data and exclusion of readers with colour-vision
deficiencies \citep{crameri_misuse_2020}.

The authors report no conflict of interest.
\end{acknowledgments}

\appendix

\section{Derivation of the boundary integral representation in an imposed
flow\label{app:Derivation-boundary-integral}}

Starting from the well-known integral representation of the Stokes
equation in the absence of any background flow \citep{ladyzhenskaya1969}
\begin{equation}
v'_{\alpha}(\boldsymbol{r})=-\int G_{\alpha\beta}(\boldsymbol{r},\boldsymbol{r}^{\prime})f_{\beta}(\boldsymbol{r}^{\prime})\,d\mathcal{S}+\int K_{\beta\alpha\nu}(\boldsymbol{r}^{\prime},\boldsymbol{r})\hat{\rho}_{\nu}^{\prime}v'_{\beta}(\boldsymbol{r}^{\prime})\,d\mathcal{S},\qquad\boldsymbol{r}\in\mathcal{V},\quad\boldsymbol{r}^{\prime}=\boldsymbol{R}+\boldsymbol{\rho}^{\prime}\in\mathcal{S},\label{eq:BI-no-imposed}
\end{equation}
we follow the derivations in \citep{pozrikidis1992,leal2007advanced}
for a situation involving an undisturbed imposed velocity field $\boldsymbol{v}^{\infty}(\boldsymbol{r})$.
In this case, $\boldsymbol{v}'$ can be interpreted as a disturbance
field due to the colloid being present in the fluid. We can thus write
$\boldsymbol{v}'=\boldsymbol{v}-\boldsymbol{v}^{\infty},$ with $\boldsymbol{v}$
now being the true velocity field. We can use the Lorentz reciprocal
theorem \citep{lorentz1907}
\begin{equation}
\boldsymbol{\nabla}\cdot\left(\boldsymbol{v}^{*}\cdot\boldsymbol{\sigma}^{\infty}-\boldsymbol{v}^{\infty}\cdot\boldsymbol{\sigma}^{*}\right)=0,\label{eq:reciprocal}
\end{equation}
for the regular imposed flow $\boldsymbol{v}^{\infty}$ and an arbitrary
regular flow $\boldsymbol{v}^{*}$, with associated stress tensors
$\boldsymbol{\sigma}^{\infty}$ and $\boldsymbol{\sigma}^{*}$, respectively,
to further simplify the result. Choosing $\boldsymbol{v}^{*}$ to
be the flow due to a Stokeslet of strength $\boldsymbol{g}$ located
at $\boldsymbol{r}$ we have the fundamental solution of the Stokes
equation
\[
\boldsymbol{v}^{*}(\boldsymbol{r}')=\boldsymbol{G}(\boldsymbol{r}',\boldsymbol{r})\cdot\boldsymbol{g},\qquad\boldsymbol{\sigma}^{*}(\boldsymbol{r}')=\boldsymbol{K}(\boldsymbol{r}',\boldsymbol{r})\cdot\boldsymbol{g}.
\]
Using this in the reciprocal theorem gives
\begin{equation}
\boldsymbol{\nabla}\cdot\left(\boldsymbol{G}(\boldsymbol{r},\boldsymbol{r}')\cdot\boldsymbol{\sigma}^{\infty}-\boldsymbol{v}^{\infty}\cdot\boldsymbol{K}(\boldsymbol{r}',\boldsymbol{r})\right)=0.\label{eq:regular}
\end{equation}
Choosing $\boldsymbol{r}$ to lie outside the (arbitrary) fluid domain
$\mathcal{V}$ and noting that the above expression in brackets is
then regular in $\mathcal{V}$, we can integrate this over $\mathcal{V}$
and use the divergence theorem to convert it into a surface integral
over the bounding surface of the chosen fluid domain, which in our
case is the surface of the colloid, to obtain
\[
\int\left(\boldsymbol{G}(\boldsymbol{r},\boldsymbol{r}')\cdot\boldsymbol{\sigma}^{\infty}-\boldsymbol{v}^{\infty}\cdot\boldsymbol{K}(\boldsymbol{r}',\boldsymbol{r})\right)\cdot\hat{\boldsymbol{\rho}}'\,d\mathcal{S}=0.
\]
Writing $\boldsymbol{f}^{\infty}=\hat{\boldsymbol{\rho}}\cdot\boldsymbol{\sigma}^{\infty}$
on the surface of the sphere, we have the identity
\[
\int\left(\boldsymbol{G}(\boldsymbol{r},\boldsymbol{r}')\cdot\boldsymbol{f}^{\infty}-\boldsymbol{v}^{\infty}\cdot\boldsymbol{K}(\boldsymbol{r}',\boldsymbol{r})\cdot\hat{\boldsymbol{\rho}}'\right)\,d\mathcal{S}=0.
\]
This yields the boundary integral representation \eqref{eq:BI},
\begin{equation}
v_{\alpha}(\boldsymbol{r})=v_{\alpha}^{\infty}(\boldsymbol{r})-\int G_{\alpha\beta}(\boldsymbol{r},\boldsymbol{r}^{\prime})f_{\beta}(\boldsymbol{r}^{\prime})\,d\mathcal{S}+\int K_{\beta\alpha\nu}(\boldsymbol{r}^{\prime},\boldsymbol{r})\hat{\rho}_{\nu}^{\prime}v_{\beta}(\boldsymbol{r}^{\prime})\,d\mathcal{S},\qquad\boldsymbol{r}\in\mathcal{V},\quad\boldsymbol{r}^{\prime}=\boldsymbol{R}+\boldsymbol{\rho}^{\prime}\in\mathcal{S}.\label{eq:BI-imposed}
\end{equation}

Here, we can also define the three contributions to the traction in
Eq. \eqref{eq:traction} as follows. Consider the boundary integral
equation, Eq. \eqref{eq:BIE}, for a rigid body with boundary condition
$\boldsymbol{v}(\boldsymbol{R}+\boldsymbol{\rho})=\boldsymbol{V}+\boldsymbol{\Omega}\times\boldsymbol{\rho}=\boldsymbol{v}^{\mathcal{D}}(\boldsymbol{\rho})$.
We use that rigid body motion is an eigenfunction of the double-layer
integral operator with eigenvalue $-1/2$ \citep{kim2015ellipsoidal}
to obtain
\[
\boldsymbol{v}^{\mathcal{D}}(\boldsymbol{r})=\boldsymbol{v}^{\infty}(\boldsymbol{r})-\int\boldsymbol{G}(\boldsymbol{r},\boldsymbol{r}^{\prime})\cdot\boldsymbol{f}(\boldsymbol{r}^{\prime})\,d\mathcal{S},
\]
If the rigid body is held stationary, i.e., $\boldsymbol{v}^{\mathcal{D}}=0,$
in the imposed flow we have 
\[
\boldsymbol{v}^{\infty}(\boldsymbol{r})=\int\boldsymbol{G}(\boldsymbol{r},\boldsymbol{r}^{\prime})\cdot\boldsymbol{f}^{\infty}(\boldsymbol{r}^{\prime})\,d\mathcal{S},
\]
which defines $\boldsymbol{f}^{\infty}$ as the traction necessary
to keep a rigid body stationary when exposed to an imposed flow $\boldsymbol{v}^{\infty}(\boldsymbol{r})$.
Using the linearity of Stokes flow, we can write for a non-stationary
rigid particle
\[
\boldsymbol{v}^{\mathcal{D}}(\boldsymbol{r})=\boldsymbol{v}^{\infty}(\boldsymbol{r})-\int\boldsymbol{G}(\boldsymbol{r},\boldsymbol{r}^{\prime})\cdot\left(\boldsymbol{f}^{\mathcal{D}}(\boldsymbol{r}^{\prime})+\boldsymbol{f}^{\infty}(\boldsymbol{r}^{\prime})\right)\,d\mathcal{S}.
\]
Let us now look at an active particle with boundary condition given
by \eqref{eq:boundary-1}. Following the same steps as above we obtain
\[
\boldsymbol{v}^{\mathcal{D}}(\boldsymbol{r})+\tfrac{1}{2}\boldsymbol{v}^{\mathcal{A}}(\boldsymbol{r})=\boldsymbol{v}^{\infty}(\boldsymbol{r})-\int\boldsymbol{G}(\boldsymbol{r},\boldsymbol{r}^{\prime})\cdot\left(\boldsymbol{f}^{\mathcal{D}}(\boldsymbol{r}^{\prime})+\boldsymbol{f}^{\infty}(\boldsymbol{r}^{\prime})+\boldsymbol{f}^{\mathcal{A}}(\boldsymbol{r}^{\prime})\right)\,d\mathcal{S}+\int\boldsymbol{v}^{\mathcal{A}}(\boldsymbol{r}^{\prime})\cdot\boldsymbol{K}(\boldsymbol{r}^{\prime},\boldsymbol{r})\cdot\hat{\boldsymbol{\rho}}'\,d\mathcal{S},
\]
with $\boldsymbol{f}^{\mathcal{A}}$ the traction caused by the active
slip. By linearity, this equation contains the three independent boundary
integral equations \eqref{eq:BIE}. As can be seen from this derivation,
the boundary integral equations for the imposed background flow $\boldsymbol{v}^{\infty}$
and activity $\boldsymbol{v}^{\mathcal{A}}$, with the present definitions
of the three distinct contributions to the traction as in \eqref{eq:traction},
cannot be written in equivalent form. This possibly unintuitive result
has been noted before in \citep{singh2018generalized}, although without
derivation.

\section{Power dissipation for uniaxial slip flow\label{app:Power-dissipation}}

With the uniaxial parametrisation introduced in the caption of Figure
\ref{fig:Slip-and-traction} we can rewrite the power dissipation
due to activity $\dot{\mathcal{E}}^{\mathcal{A}}$ in the following
way. We have\begin{subequations}
\begin{align}
\boldsymbol{V}^{\mathcal{A}(ls)}\odot\boldsymbol{V}^{\mathcal{A}(ls)} & =\left(V_{ls}^{0,\mathcal{A}}\right)^{2}\boldsymbol{Y}^{(l)}(\boldsymbol{e})\odot\boldsymbol{Y}^{(l)}(\boldsymbol{e}),\\
\boldsymbol{V}^{\mathcal{A}(la)}\odot\boldsymbol{V}^{\mathcal{A}(la)} & =\left(V_{la}^{0,\mathcal{A}}\right)^{2}\boldsymbol{Y}^{(l-1)}(\boldsymbol{e})\odot\boldsymbol{Y}^{(l-1)}(\boldsymbol{e}),\\
\boldsymbol{V}^{\mathcal{A}(lt)}\odot\boldsymbol{V}^{\mathcal{A}(lt)} & =\left(V_{lt}^{0,\mathcal{A}}\right)^{2}\boldsymbol{Y}^{(l-2)}(\boldsymbol{e})\odot\boldsymbol{Y}^{(l-2)}(\boldsymbol{e}),
\end{align}
\end{subequations}and with the orthogonality relation of TSH \eqref{eq:orthoTSH}
and the identity $\Delta_{\mu_{1}\dots\mu_{l},\mu_{1}\dots\mu_{l}}^{(l)}=2l+1$
\citep{brunn1976effect} one can show that
\begin{equation}
\boldsymbol{Y}^{(l)}(\boldsymbol{e})\odot\boldsymbol{Y}^{(l)}(\boldsymbol{e})=\frac{1}{w_{l+1}}.
\end{equation}
Thus the power dissipation in terms of the friction coefficients and
the strengths of the slip modes is
\begin{gather}
\dot{\mathcal{E}}^{\mathcal{A}}=\frac{\hat{\gamma}_{ls}}{w_{l+1}}\,\left(V_{ls}^{0,\mathcal{A}}\right)^{2}+\frac{2}{2l-3}\left(\frac{l-1}{l}\right)^{2}\frac{\hat{\gamma}_{la}}{w_{l}}\,\left(V_{la}^{0,\mathcal{A}}\right)^{2}+\frac{2l-3}{2l-1}\frac{\hat{\gamma}_{lt}}{w_{l-1}}\,\left(V_{lt}^{0,\mathcal{A}}\right)^{2}\label{eq:powDissAlt}
\end{gather}
implicitly summing over all slip modes that are present. With this
we can compute the power dissipated by any isolated mode of slip,
which is potentially useful in optimisation problems such as the question
for the most efficient way to swim for a certain microorganism \citep{lighthill1952,daddi-moussa-ider_optimal_2021,guo_optimal_2021}.


\begin{thebibliography}{89}%
\makeatletter
\providecommand \@ifxundefined [1]{%
 \@ifx{#1\undefined}
}%
\providecommand \@ifnum [1]{%
 \ifnum #1\expandafter \@firstoftwo
 \else \expandafter \@secondoftwo
 \fi
}%
\providecommand \@ifx [1]{%
 \ifx #1\expandafter \@firstoftwo
 \else \expandafter \@secondoftwo
 \fi
}%
\providecommand \natexlab [1]{#1}%
\providecommand \enquote  [1]{``#1''}%
\providecommand \bibnamefont  [1]{#1}%
\providecommand \bibfnamefont [1]{#1}%
\providecommand \citenamefont [1]{#1}%
\providecommand \href@noop [0]{\@secondoftwo}%
\providecommand \href [0]{\begingroup \@sanitize@url \@href}%
\providecommand \@href[1]{\@@startlink{#1}\@@href}%
\providecommand \@@href[1]{\endgroup#1\@@endlink}%
\providecommand \@sanitize@url [0]{\catcode `\\12\catcode `\$12\catcode
  `\&12\catcode `\#12\catcode `\^12\catcode `\_12\catcode `\%12\relax}%
\providecommand \@@startlink[1]{}%
\providecommand \@@endlink[0]{}%
\providecommand \url  [0]{\begingroup\@sanitize@url \@url }%
\providecommand \@url [1]{\endgroup\@href {#1}{\urlprefix }}%
\providecommand \urlprefix  [0]{URL }%
\providecommand \Eprint [0]{\href }%
\providecommand \doibase [0]{http://dx.doi.org/}%
\providecommand \selectlanguage [0]{\@gobble}%
\providecommand \bibinfo  [0]{\@secondoftwo}%
\providecommand \bibfield  [0]{\@secondoftwo}%
\providecommand \translation [1]{[#1]}%
\providecommand \BibitemOpen [0]{}%
\providecommand \bibitemStop [0]{}%
\providecommand \bibitemNoStop [0]{.\EOS\space}%
\providecommand \EOS [0]{\spacefactor3000\relax}%
\providecommand \BibitemShut  [1]{\csname bibitem#1\endcsname}%
\let\auto@bib@innerbib\@empty
\bibitem [{\citenamefont {Paxton}\ \emph {et~al.}(2004)\citenamefont {Paxton},
  \citenamefont {C.}, \citenamefont {Olmeda}, \citenamefont {Sen},
  \citenamefont {Angelo}, \citenamefont {Cao}, \citenamefont {Mallouk},
  \citenamefont {Lammert},\ and\ \citenamefont {Crespi}}]{paxton2004}%
  \BibitemOpen
  \bibfield  {author} {\bibinfo {author} {\bibfnamefont {W.~F.}\ \bibnamefont
  {Paxton}}, \bibinfo {author} {\bibfnamefont {K.~C.~Kevin}\ \bibnamefont
  {C.}}, \bibinfo {author} {\bibfnamefont {C.~C.}\ \bibnamefont {Olmeda}},
  \bibinfo {author} {\bibfnamefont {A.}~\bibnamefont {Sen}}, \bibinfo {author}
  {\bibfnamefont {S.~K.~St.}\ \bibnamefont {Angelo}}, \bibinfo {author}
  {\bibfnamefont {Y.}~\bibnamefont {Cao}}, \bibinfo {author} {\bibfnamefont
  {T.~E.}\ \bibnamefont {Mallouk}}, \bibinfo {author} {\bibfnamefont {P.~E.}\
  \bibnamefont {Lammert}}, \ and\ \bibinfo {author} {\bibfnamefont {V.~H.}\
  \bibnamefont {Crespi}},\ }\bibfield  {title} {\enquote {\bibinfo {title}
  {{Catalytic nanomotors: Autonomous movement of striped nanorods}},}\ }\href
  {\doibase 10.1021/ja047697z} {\bibfield  {journal} {\bibinfo  {journal} {J.
  Am. Chem. Soc.}\ }\textbf {\bibinfo {volume} {126}},\ \bibinfo {pages}
  {13424--13431} (\bibinfo {year} {2004})}\BibitemShut {NoStop}%
\bibitem [{\citenamefont {Howse}\ \emph {et~al.}(2007)\citenamefont {Howse},
  \citenamefont {Jones}, \citenamefont {Ryan}, \citenamefont {Gough},
  \citenamefont {Vafabakhsh},\ and\ \citenamefont
  {Golestanian}}]{howse2007self}%
  \BibitemOpen
  \bibfield  {author} {\bibinfo {author} {\bibfnamefont {J.~R.}\ \bibnamefont
  {Howse}}, \bibinfo {author} {\bibfnamefont {R.~A.~L.}\ \bibnamefont {Jones}},
  \bibinfo {author} {\bibfnamefont {A.~J.}\ \bibnamefont {Ryan}}, \bibinfo
  {author} {\bibfnamefont {T.}~\bibnamefont {Gough}}, \bibinfo {author}
  {\bibfnamefont {R.}~\bibnamefont {Vafabakhsh}}, \ and\ \bibinfo {author}
  {\bibfnamefont {R.}~\bibnamefont {Golestanian}},\ }\bibfield  {title}
  {\enquote {\bibinfo {title} {Self-motile colloidal particles: {F}rom directed
  propulsion to random walk},}\ }\href {\doibase 10.1103/PhysRevLett.99.048102}
  {\bibfield  {journal} {\bibinfo  {journal} {Phys. Rev. Lett.}\ }\textbf
  {\bibinfo {volume} {99}},\ \bibinfo {pages} {048102} (\bibinfo {year}
  {2007})}\BibitemShut {NoStop}%
\bibitem [{\citenamefont {Jiang}\ \emph {et~al.}(2010)\citenamefont {Jiang},
  \citenamefont {Yoshinaga},\ and\ \citenamefont {Sano}}]{jiang2010active}%
  \BibitemOpen
  \bibfield  {author} {\bibinfo {author} {\bibfnamefont {H.-R.}\ \bibnamefont
  {Jiang}}, \bibinfo {author} {\bibfnamefont {N.}~\bibnamefont {Yoshinaga}}, \
  and\ \bibinfo {author} {\bibfnamefont {M.}~\bibnamefont {Sano}},\ }\bibfield
  {title} {\enquote {\bibinfo {title} {Active motion of a janus particle by
  self-thermophoresis in a defocused laser beam},}\ }\href {\doibase
  10.1103/PhysRevLett.105.268302} {\bibfield  {journal} {\bibinfo  {journal}
  {Phys. Rev. Lett.}\ }\textbf {\bibinfo {volume} {105}},\ \bibinfo {pages}
  {268302} (\bibinfo {year} {2010})}\BibitemShut {NoStop}%
\bibitem [{\citenamefont {Ebbens}\ and\ \citenamefont
  {Howse}(2010)}]{ebbens2010pursuit}%
  \BibitemOpen
  \bibfield  {author} {\bibinfo {author} {\bibfnamefont {S.~J.}\ \bibnamefont
  {Ebbens}}\ and\ \bibinfo {author} {\bibfnamefont {J.~R.}\ \bibnamefont
  {Howse}},\ }\bibfield  {title} {\enquote {\bibinfo {title} {In pursuit of
  propulsion at the nanoscale},}\ }\href {\doibase 10.1039/B918598D} {\bibfield
   {journal} {\bibinfo  {journal} {Soft Matter}\ }\textbf {\bibinfo {volume}
  {6}},\ \bibinfo {pages} {726--738} (\bibinfo {year} {2010})}\BibitemShut
  {NoStop}%
\bibitem [{\citenamefont {Palacci}\ \emph {et~al.}(2013)\citenamefont
  {Palacci}, \citenamefont {Sacanna}, \citenamefont {Steinberg}, \citenamefont
  {Pine},\ and\ \citenamefont {Chaikin}}]{palacci2013living}%
  \BibitemOpen
  \bibfield  {author} {\bibinfo {author} {\bibfnamefont {J.}~\bibnamefont
  {Palacci}}, \bibinfo {author} {\bibfnamefont {S.}~\bibnamefont {Sacanna}},
  \bibinfo {author} {\bibfnamefont {A.~P.}\ \bibnamefont {Steinberg}}, \bibinfo
  {author} {\bibfnamefont {D.~J.}\ \bibnamefont {Pine}}, \ and\ \bibinfo
  {author} {\bibfnamefont {P.~M.}\ \bibnamefont {Chaikin}},\ }\bibfield
  {title} {\enquote {\bibinfo {title} {Living crystals of light-activated
  colloidal surfers},}\ }\href {\doibase 10.1126/science.1230020} {\bibfield
  {journal} {\bibinfo  {journal} {Science}\ }\textbf {\bibinfo {volume}
  {339}},\ \bibinfo {pages} {936--940} (\bibinfo {year} {2013})}\BibitemShut
  {NoStop}%
\bibitem [{\citenamefont {Brennen}\ and\ \citenamefont
  {Winet}(1977)}]{brennen1977}%
  \BibitemOpen
  \bibfield  {author} {\bibinfo {author} {\bibfnamefont {C.}~\bibnamefont
  {Brennen}}\ and\ \bibinfo {author} {\bibfnamefont {H.}~\bibnamefont
  {Winet}},\ }\bibfield  {title} {\enquote {\bibinfo {title} {{Fluid mechanics
  of propulsion by cilia and flagella}},}\ }\href {\doibase
  10.1146/annurev.fl.09.010177.002011} {\bibfield  {journal} {\bibinfo
  {journal} {Annu. Rev. Fluid Mech.}\ }\textbf {\bibinfo {volume} {9}},\
  \bibinfo {pages} {339--398} (\bibinfo {year} {1977})}\BibitemShut {NoStop}%
\bibitem [{\citenamefont {Lighthill}(1952)}]{lighthill1952}%
  \BibitemOpen
  \bibfield  {author} {\bibinfo {author} {\bibfnamefont {M.~J.}\ \bibnamefont
  {Lighthill}},\ }\bibfield  {title} {\enquote {\bibinfo {title} {On the
  squirming motion of nearly spherical deformable bodies through liquids at
  very small {R}eynolds numbers},}\ }\href {\doibase 10.1002/cpa.3160050201}
  {\bibfield  {journal} {\bibinfo  {journal} {Commun. Pure. Appl. Math.}\
  }\textbf {\bibinfo {volume} {5}},\ \bibinfo {pages} {109--118} (\bibinfo
  {year} {1952})}\BibitemShut {NoStop}%
\bibitem [{\citenamefont {Blake}(1971{\natexlab{a}})}]{blake1971a}%
  \BibitemOpen
  \bibfield  {author} {\bibinfo {author} {\bibfnamefont {J.~R.}\ \bibnamefont
  {Blake}},\ }\bibfield  {title} {\enquote {\bibinfo {title} {{A spherical
  envelope approach to ciliary propulsion}},}\ }\href {\doibase
  10.1017/S002211207100048X} {\bibfield  {journal} {\bibinfo  {journal} {J.
  Fluid Mech.}\ }\textbf {\bibinfo {volume} {46}},\ \bibinfo {pages} {199--208}
  (\bibinfo {year} {1971}{\natexlab{a}})}\BibitemShut {NoStop}%
\bibitem [{\citenamefont {Anderson}(1989)}]{anderson1989colloid}%
  \BibitemOpen
  \bibfield  {author} {\bibinfo {author} {\bibfnamefont {J.~L.}\ \bibnamefont
  {Anderson}},\ }\bibfield  {title} {\enquote {\bibinfo {title} {Colloid
  transport by interfacial forces},}\ }\href {\doibase
  10.1146/annurev.fl.21.010189.000425} {\bibfield  {journal} {\bibinfo
  {journal} {Annu. Rev. Fluid Mech.}\ }\textbf {\bibinfo {volume} {21}},\
  \bibinfo {pages} {61--99} (\bibinfo {year} {1989})}\BibitemShut {NoStop}%
\bibitem [{\citenamefont {Stokes}(1850)}]{stokesEffectInternalFriction1850}%
  \BibitemOpen
  \bibfield  {author} {\bibinfo {author} {\bibfnamefont {George~Gabriel}\
  \bibnamefont {Stokes}},\ }\href {\doibase 10.1017/CBO9780511702266} {\emph
  {\bibinfo {title} {On the effect of the internal friction of fluids on the
  motion of pendulums}}}\ (\bibinfo  {publisher} {Cambridge University Press},\
  \bibinfo {address} {Cambridge},\ \bibinfo {year} {1850})\BibitemShut
  {NoStop}%
\bibitem [{\citenamefont {Landau}\ and\ \citenamefont
  {Lifshitz}(1959)}]{landau1959fluid}%
  \BibitemOpen
  \bibfield  {author} {\bibinfo {author} {\bibfnamefont {L.~D.}\ \bibnamefont
  {Landau}}\ and\ \bibinfo {author} {\bibfnamefont {E.~M.}\ \bibnamefont
  {Lifshitz}},\ }\href {https://archive.org/details/FluidMechanics} {\emph
  {\bibinfo {title} {Fluid mechanics}}},\ Vol.~\bibinfo {volume} {6}\ (\bibinfo
   {publisher} {Pergamon Press, New York},\ \bibinfo {year} {1959})\BibitemShut
  {NoStop}%
\bibitem [{\citenamefont {Brunn}(1976)}]{brunn1976effect}%
  \BibitemOpen
  \bibfield  {author} {\bibinfo {author} {\bibfnamefont {P.}~\bibnamefont
  {Brunn}},\ }\bibfield  {title} {\enquote {\bibinfo {title} {The effect of
  {B}rownian motion for a suspension of spheres},}\ }\href {\doibase
  10.1007/BF01517501} {\bibfield  {journal} {\bibinfo  {journal} {Rheol. Acta}\
  }\textbf {\bibinfo {volume} {15}},\ \bibinfo {pages} {104--119} (\bibinfo
  {year} {1976})}\BibitemShut {NoStop}%
\bibitem [{\citenamefont {Brunn}(1979)}]{brunnMotionSlightlyDeformed1979}%
  \BibitemOpen
  \bibfield  {author} {\bibinfo {author} {\bibfnamefont {P.}~\bibnamefont
  {Brunn}},\ }\bibfield  {title} {\enquote {\bibinfo {title} {The motion of a
  slightly deformed sphere in a viscoelastic fluid},}\ }\href {\doibase
  10.1007/BF01542770} {\bibfield  {journal} {\bibinfo  {journal} {Rheologica
  Acta}\ }\textbf {\bibinfo {volume} {18}},\ \bibinfo {pages} {229--243}
  (\bibinfo {year} {1979})}\BibitemShut {NoStop}%
\bibitem [{\citenamefont {Pak}\ and\ \citenamefont
  {Lauga}(2014)}]{pak2014generalized}%
  \BibitemOpen
  \bibfield  {author} {\bibinfo {author} {\bibfnamefont {O.~S.}\ \bibnamefont
  {Pak}}\ and\ \bibinfo {author} {\bibfnamefont {E.}~\bibnamefont {Lauga}},\
  }\bibfield  {title} {\enquote {\bibinfo {title} {Generalized squirming motion
  of a sphere},}\ }\href {\doibase 10.1007/s10665-014-9690-9} {\bibfield
  {journal} {\bibinfo  {journal} {J. Eng. Math.}\ }\textbf {\bibinfo {volume}
  {88}},\ \bibinfo {pages} {1--28} (\bibinfo {year} {2014})}\BibitemShut
  {NoStop}%
\bibitem [{\citenamefont {Pedley}(2016)}]{pedley2016spherical}%
  \BibitemOpen
  \bibfield  {author} {\bibinfo {author} {\bibfnamefont {T.~J.}\ \bibnamefont
  {Pedley}},\ }\bibfield  {title} {\enquote {\bibinfo {title} {Spherical
  squirmers: models for swimming micro-organisms},}\ }\href {\doibase
  10.1093/imamat/hxw030} {\bibfield  {journal} {\bibinfo  {journal} {IMA J.
  Appl. Math.}\ }\textbf {\bibinfo {volume} {81}},\ \bibinfo {pages} {488--521}
  (\bibinfo {year} {2016})}\BibitemShut {NoStop}%
\bibitem [{\citenamefont {Pedley}\ \emph {et~al.}(2016)\citenamefont {Pedley},
  \citenamefont {Brumley},\ and\ \citenamefont
  {Goldstein}}]{pedley_brumley_goldstein_2016}%
  \BibitemOpen
  \bibfield  {author} {\bibinfo {author} {\bibfnamefont {T.~J.}\ \bibnamefont
  {Pedley}}, \bibinfo {author} {\bibfnamefont {D.~R.}\ \bibnamefont {Brumley}},
  \ and\ \bibinfo {author} {\bibfnamefont {R.~E.}\ \bibnamefont {Goldstein}},\
  }\bibfield  {title} {\enquote {\bibinfo {title} {Squirmers with swirl: a
  model for volvox swimming},}\ }\href {\doibase 10.1017/jfm.2016.306}
  {\bibfield  {journal} {\bibinfo  {journal} {J. Fluid Mech.}\ }\textbf
  {\bibinfo {volume} {798}},\ \bibinfo {pages} {165--186} (\bibinfo {year}
  {2016})}\BibitemShut {NoStop}%
\bibitem [{\citenamefont {Rojas-P{\'e}rez}\ \emph {et~al.}(2021)\citenamefont
  {Rojas-P{\'e}rez}, \citenamefont {Delmotte},\ and\ \citenamefont
  {Michelin}}]{rojas2021hydrochemical}%
  \BibitemOpen
  \bibfield  {author} {\bibinfo {author} {\bibfnamefont {F.}~\bibnamefont
  {Rojas-P{\'e}rez}}, \bibinfo {author} {\bibfnamefont {B.}~\bibnamefont
  {Delmotte}}, \ and\ \bibinfo {author} {\bibfnamefont {S.}~\bibnamefont
  {Michelin}},\ }\bibfield  {title} {\enquote {\bibinfo {title} {Hydrochemical
  interactions of phoretic particles: a regularized multipole framework},}\
  }\href@noop {} {\bibfield  {journal} {\bibinfo  {journal} {J. Fluid Mech.}\
  }\textbf {\bibinfo {volume} {919}},\ \bibinfo {pages} {A22} (\bibinfo {year}
  {2021})}\BibitemShut {NoStop}%
\bibitem [{\citenamefont {Lorentz}(1907)}]{lorentz1907}%
  \BibitemOpen
  \bibfield  {author} {\bibinfo {author} {\bibfnamefont {H.~A.}\ \bibnamefont
  {Lorentz}},\ }\bibfield  {title} {\enquote {\bibinfo {title} {A general
  theory concerning the motion of a viscous fluid},}\ }\href@noop {} {\bibfield
   {journal} {\bibinfo  {journal} {Abhandl. Theor. Phys}\ }\textbf {\bibinfo
  {volume} {1}},\ \bibinfo {pages} {23--42} (\bibinfo {year}
  {1907})}\BibitemShut {NoStop}%
\bibitem [{\citenamefont {Odqvist}(1930)}]{fkg1930bandwertaufgaben}%
  \BibitemOpen
  \bibfield  {author} {\bibinfo {author} {\bibfnamefont {F.~K.~G.}\
  \bibnamefont {Odqvist}},\ }\bibfield  {title} {\enquote {\bibinfo {title}
  {{\"U}ber die bandwertaufgaben der hydrodynamik z{\"a}her
  fl{\"u}ssig-keiten},}\ }\href {\doibase 10.1007/BF01194638} {\bibfield
  {journal} {\bibinfo  {journal} {Mathematische Zeitschrift}\ }\textbf
  {\bibinfo {volume} {32}},\ \bibinfo {pages} {329--375} (\bibinfo {year}
  {1930})}\BibitemShut {NoStop}%
\bibitem [{\citenamefont {Ladyzhenskaia}(1969)}]{ladyzhenskaya1969}%
  \BibitemOpen
  \bibfield  {author} {\bibinfo {author} {\bibfnamefont {O.~A.}\ \bibnamefont
  {Ladyzhenskaia}},\ }\href {https://books.google.co.in/books?id=qVXvAAAAMAAJ}
  {\emph {\bibinfo {title} {The mathematical theory of viscous incompressible
  flow}}},\ Mathematics and its applications\ (\bibinfo  {publisher} {Gordon
  and Breach},\ \bibinfo {year} {1969})\BibitemShut {NoStop}%
\bibitem [{\citenamefont {Cheng}\ and\ \citenamefont
  {Cheng}(2005)}]{cheng2005heritage}%
  \BibitemOpen
  \bibfield  {author} {\bibinfo {author} {\bibfnamefont {A.~H.-D.}\
  \bibnamefont {Cheng}}\ and\ \bibinfo {author} {\bibfnamefont {D.~T.}\
  \bibnamefont {Cheng}},\ }\bibfield  {title} {\enquote {\bibinfo {title}
  {Heritage and early history of the boundary element method},}\ }\href
  {\doibase 10.1016/j.enganabound.2004.12.001} {\bibfield  {journal} {\bibinfo
  {journal} {Eng. Anal. Bound. Elem.}\ }\textbf {\bibinfo {volume} {29}},\
  \bibinfo {pages} {268--302} (\bibinfo {year} {2005})}\BibitemShut {NoStop}%
\bibitem [{\citenamefont {Youngren}\ and\ \citenamefont
  {Acrivos}(1975)}]{youngren1975stokes}%
  \BibitemOpen
  \bibfield  {author} {\bibinfo {author} {\bibfnamefont {G.}~\bibnamefont
  {Youngren}}\ and\ \bibinfo {author} {\bibfnamefont {A.}~\bibnamefont
  {Acrivos}},\ }\bibfield  {title} {\enquote {\bibinfo {title} {Stokes flow
  past a particle of arbitrary shape: a numerical method of solution},}\ }\href
  {\doibase 10.1017/S0022112075001486} {\bibfield  {journal} {\bibinfo
  {journal} {J. Fluid Mech.}\ }\textbf {\bibinfo {volume} {69}},\ \bibinfo
  {pages} {377--403} (\bibinfo {year} {1975})}\BibitemShut {NoStop}%
\bibitem [{\citenamefont {Felderhof}(1976{\natexlab{a}})}]{felderhof1976}%
  \BibitemOpen
  \bibfield  {author} {\bibinfo {author} {\bibfnamefont {B.~U.}\ \bibnamefont
  {Felderhof}},\ }\bibfield  {title} {\enquote {\bibinfo {title} {Force density
  induced on a sphere in linear hydrodynamics: I. {F}ixed sphere, stick
  boundary conditions},}\ }\href {\doibase 10.1016/0378-4371(76)90104-7}
  {\bibfield  {journal} {\bibinfo  {journal} {Physica A}\ }\textbf {\bibinfo
  {volume} {84}},\ \bibinfo {pages} {557--568} (\bibinfo {year}
  {1976}{\natexlab{a}})}\BibitemShut {NoStop}%
\bibitem [{\citenamefont
  {Felderhof}(1976{\natexlab{b}})}]{felderhofForceDensityInduced1976a}%
  \BibitemOpen
  \bibfield  {author} {\bibinfo {author} {\bibfnamefont {B.~U.}\ \bibnamefont
  {Felderhof}},\ }\bibfield  {title} {\enquote {\bibinfo {title} {Force density
  induced on a sphere in linear hydrodynamics: {II}. {Moving} sphere, mixed
  boundary conditions},}\ }\href {\doibase 10.1016/0378-4371(76)90105-9}
  {\bibfield  {journal} {\bibinfo  {journal} {Physica A: Statistical Mechanics
  and its Applications}\ }\textbf {\bibinfo {volume} {84}},\ \bibinfo {pages}
  {569--576} (\bibinfo {year} {1976}{\natexlab{b}})}\BibitemShut {NoStop}%
\bibitem [{\citenamefont {Zick}\ and\ \citenamefont
  {Homsy}(1982)}]{zick1982stokes}%
  \BibitemOpen
  \bibfield  {author} {\bibinfo {author} {\bibfnamefont {A.~A.}\ \bibnamefont
  {Zick}}\ and\ \bibinfo {author} {\bibfnamefont {G.~M.}\ \bibnamefont
  {Homsy}},\ }\bibfield  {title} {\enquote {\bibinfo {title} {Stokes flow
  through periodic arrays of spheres},}\ }\href {\doibase
  10.1017/S0022112082000627} {\bibfield  {journal} {\bibinfo  {journal} {J.
  Fluid Mech.}\ }\textbf {\bibinfo {volume} {115}},\ \bibinfo {pages} {13--26}
  (\bibinfo {year} {1982})}\BibitemShut {NoStop}%
\bibitem [{\citenamefont {Schmitz}\ and\ \citenamefont
  {Felderhof}(1982)}]{schmitzCreepingFlowSpherical1982}%
  \BibitemOpen
  \bibfield  {author} {\bibinfo {author} {\bibfnamefont {R.}~\bibnamefont
  {Schmitz}}\ and\ \bibinfo {author} {\bibfnamefont {B.~U.}\ \bibnamefont
  {Felderhof}},\ }\bibfield  {title} {\enquote {\bibinfo {title} {Creeping flow
  about a spherical particle},}\ }\href {\doibase 10.1016/0378-4371(82)90007-3}
  {\bibfield  {journal} {\bibinfo  {journal} {Physica A: Statistical Mechanics
  and its Applications}\ }\textbf {\bibinfo {volume} {113}},\ \bibinfo {pages}
  {90--102} (\bibinfo {year} {1982})}\BibitemShut {NoStop}%
\bibitem [{\citenamefont {Ghose}\ and\ \citenamefont
  {Adhikari}(2014)}]{ghose2014irreducible}%
  \BibitemOpen
  \bibfield  {author} {\bibinfo {author} {\bibfnamefont {S.}~\bibnamefont
  {Ghose}}\ and\ \bibinfo {author} {\bibfnamefont {R.}~\bibnamefont
  {Adhikari}},\ }\bibfield  {title} {\enquote {\bibinfo {title} {Irreducible
  representations of oscillatory and swirling flows in active soft matter},}\
  }\href {\doibase 10.1103/PhysRevLett.112.118102} {\bibfield  {journal}
  {\bibinfo  {journal} {Phys. Rev. Lett.}\ }\textbf {\bibinfo {volume} {112}},\
  \bibinfo {pages} {118102} (\bibinfo {year} {2014})}\BibitemShut {NoStop}%
\bibitem [{\citenamefont {Singh}\ \emph {et~al.}(2015)\citenamefont {Singh},
  \citenamefont {Ghose},\ and\ \citenamefont {Adhikari}}]{singh2015many}%
  \BibitemOpen
  \bibfield  {author} {\bibinfo {author} {\bibfnamefont {R.}~\bibnamefont
  {Singh}}, \bibinfo {author} {\bibfnamefont {S.}~\bibnamefont {Ghose}}, \ and\
  \bibinfo {author} {\bibfnamefont {R.}~\bibnamefont {Adhikari}},\ }\bibfield
  {title} {\enquote {\bibinfo {title} {Many-body microhydrodynamics of
  colloidal particles with active boundary layers},}\ }\href {\doibase
  10.1088/1742-5468/2015/06/p06017} {\bibfield  {journal} {\bibinfo  {journal}
  {J. Stat. Mech}\ }\textbf {\bibinfo {volume} {2015}},\ \bibinfo {pages}
  {P06017} (\bibinfo {year} {2015})}\BibitemShut {NoStop}%
\bibitem [{\citenamefont {Singh}\ and\ \citenamefont
  {Adhikari}(2017)}]{singh2017fluctuation}%
  \BibitemOpen
  \bibfield  {author} {\bibinfo {author} {\bibfnamefont {R.}~\bibnamefont
  {Singh}}\ and\ \bibinfo {author} {\bibfnamefont {R.}~\bibnamefont
  {Adhikari}},\ }\bibfield  {title} {\enquote {\bibinfo {title} {Fluctuating
  hydrodynamics and the {B}rownian motion of an active colloid near a wall},}\
  }\href {\doibase 10.1080/17797179.2017.1294829} {\bibfield  {journal}
  {\bibinfo  {journal} {Eur. J. Comp. Mech}\ }\textbf {\bibinfo {volume}
  {26}},\ \bibinfo {pages} {78--97} (\bibinfo {year} {2017})}\BibitemShut
  {NoStop}%
\bibitem [{\citenamefont {Singh}\ and\ \citenamefont
  {Adhikari}(2018)}]{singh2018generalized}%
  \BibitemOpen
  \bibfield  {author} {\bibinfo {author} {\bibfnamefont {R.}~\bibnamefont
  {Singh}}\ and\ \bibinfo {author} {\bibfnamefont {R.}~\bibnamefont
  {Adhikari}},\ }\bibfield  {title} {\enquote {\bibinfo {title} {Generalized
  {S}tokes laws for active colloids and their applications},}\ }\href {\doibase
  10.1088/2399-6528/aaab0d} {\bibfield  {journal} {\bibinfo  {journal} {J.
  Phys. Commun.}\ }\textbf {\bibinfo {volume} {2}},\ \bibinfo {pages} {025025}
  (\bibinfo {year} {2018})}\BibitemShut {NoStop}%
\bibitem [{\citenamefont {Boyd}(2000)}]{boyd2001chebyshev}%
  \BibitemOpen
  \bibfield  {author} {\bibinfo {author} {\bibfnamefont {J.~P.}\ \bibnamefont
  {Boyd}},\ }\href {https://store.doverpublications.com/0486411834.html} {\emph
  {\bibinfo {title} {Chebyshev and Fourier spectral methods}}}\ (\bibinfo
  {publisher} {Dover},\ \bibinfo {year} {2000})\BibitemShut {NoStop}%
\bibitem [{\citenamefont {Finlayson}\ and\ \citenamefont
  {Scriven}(1966)}]{finlayson1966method}%
  \BibitemOpen
  \bibfield  {author} {\bibinfo {author} {\bibfnamefont {B.~A.}\ \bibnamefont
  {Finlayson}}\ and\ \bibinfo {author} {\bibfnamefont {L.~E.}\ \bibnamefont
  {Scriven}},\ }\bibfield  {title} {\enquote {\bibinfo {title} {The method of
  weighted residuals - a review},}\ }\href
  {http://faculty.washington.edu/finlayso/MWR-AReview.pdf} {\bibfield
  {journal} {\bibinfo  {journal} {Appl. Mech. Rev}\ }\textbf {\bibinfo {volume}
  {19}},\ \bibinfo {pages} {735--748} (\bibinfo {year} {1966})}\BibitemShut
  {NoStop}%
\bibitem [{\citenamefont {Happel}\ and\ \citenamefont
  {Brenner}(1965)}]{happel1965low}%
  \BibitemOpen
  \bibfield  {author} {\bibinfo {author} {\bibfnamefont {J.}~\bibnamefont
  {Happel}}\ and\ \bibinfo {author} {\bibfnamefont {H.}~\bibnamefont
  {Brenner}},\ }\href
  {https://books.google.co.in/books/about/Low_Reynolds_number_hydrodynamics.html?id=_4MpAQAAMAAJ&redir_esc=y}
  {\emph {\bibinfo {title} {Low {R}eynolds number hydrodynamics: with special
  applications to particulate media}}},\ Vol.~\bibinfo {volume} {1}\ (\bibinfo
  {publisher} {Prentice-Hall},\ \bibinfo {year} {1965})\BibitemShut {NoStop}%
\bibitem [{\citenamefont {Brunn}(1980)}]{brunn1980faxen}%
  \BibitemOpen
  \bibfield  {author} {\bibinfo {author} {\bibfnamefont {P.}~\bibnamefont
  {Brunn}},\ }\bibfield  {title} {\enquote {\bibinfo {title} {Fax{\'e}n
  relations of arbitrary order and their application},}\ }\href@noop {}
  {\bibfield  {journal} {\bibinfo  {journal} {Zeitschrift f{\"u}r angewandte
  Mathematik und Physik ZAMP}\ }\textbf {\bibinfo {volume} {31}},\ \bibinfo
  {pages} {332--343} (\bibinfo {year} {1980})}\BibitemShut {NoStop}%
\bibitem [{\citenamefont {Hess}(2015)}]{hess2015tensors}%
  \BibitemOpen
  \bibfield  {author} {\bibinfo {author} {\bibfnamefont {S.}~\bibnamefont
  {Hess}},\ }\href {\doibase 10.1007/978-3-319-12787-3} {\emph {\bibinfo
  {title} {Tensors for physics}}}\ (\bibinfo  {publisher} {Springer},\ \bibinfo
  {year} {2015})\BibitemShut {NoStop}%
\bibitem [{\citenamefont {Fax{\'e}n}(1922)}]{faxen1922widerstand}%
  \BibitemOpen
  \bibfield  {author} {\bibinfo {author} {\bibfnamefont {H.}~\bibnamefont
  {Fax{\'e}n}},\ }\bibfield  {title} {\enquote {\bibinfo {title} {Der
  widerstand gegen die bewegung einer starren kugel in einer z{\"a}hen
  fl{\"u}ssigkeit, die zwischen zwei parallelen ebenen w{\"a}nden
  eingeschlossen ist},}\ }\href {\doibase 10.1002/andp.19223731003} {\bibfield
  {journal} {\bibinfo  {journal} {Ann. der Physik}\ }\textbf {\bibinfo {volume}
  {373}},\ \bibinfo {pages} {89--119} (\bibinfo {year} {1922})}\BibitemShut
  {NoStop}%
\bibitem [{\citenamefont {Batchelor}\ and\ \citenamefont
  {Green}(1972)}]{batchelorGreen1972}%
  \BibitemOpen
  \bibfield  {author} {\bibinfo {author} {\bibfnamefont {G.~K.}\ \bibnamefont
  {Batchelor}}\ and\ \bibinfo {author} {\bibfnamefont {J.~T.}\ \bibnamefont
  {Green}},\ }\bibfield  {title} {\enquote {\bibinfo {title} {The hydrodynamic
  interaction of two small freely-moving spheres in a linear flow field},}\
  }\href {\doibase 10.1017/S0022112072002927} {\bibfield  {journal} {\bibinfo
  {journal} {J. Fluid Mech.}\ }\textbf {\bibinfo {volume} {56}},\ \bibinfo
  {pages} {375--400} (\bibinfo {year} {1972})}\BibitemShut {NoStop}%
\bibitem [{\citenamefont {Rallison}(1978)}]{rallison1978note}%
  \BibitemOpen
  \bibfield  {author} {\bibinfo {author} {\bibfnamefont {J.~M.}\ \bibnamefont
  {Rallison}},\ }\bibfield  {title} {\enquote {\bibinfo {title} {Note on the
  {F}ax{\'e}n relations for a particle in {S}tokes flow},}\ }\href {\doibase
  10.1017/S0022112078002256} {\bibfield  {journal} {\bibinfo  {journal} {J.
  Fluid Mech.}\ }\textbf {\bibinfo {volume} {88}},\ \bibinfo {pages} {529--533}
  (\bibinfo {year} {1978})}\BibitemShut {NoStop}%
\bibitem [{\citenamefont
  {Barrat}(1999)}]{barratInfluenceWettingProperties1999}%
  \BibitemOpen
  \bibfield  {author} {\bibinfo {author} {\bibfnamefont {Jean-Louis}\
  \bibnamefont {Barrat}},\ }\bibfield  {title} {\enquote {\bibinfo {title}
  {Influence of wetting properties on hydrodynamic boundary conditions at a
  fluid/solid interface},}\ }\href@noop {} {\bibfield  {journal} {\bibinfo
  {journal} {Faraday Discuss.}\ }\textbf {\bibinfo {volume} {112}},\ \bibinfo
  {pages} {119--128} (\bibinfo {year} {1999})}\BibitemShut {NoStop}%
\bibitem [{\citenamefont {Lauga}\ and\ \citenamefont
  {Squires}(2005)}]{laugaBrownianMotionPartialslip2005}%
  \BibitemOpen
  \bibfield  {author} {\bibinfo {author} {\bibfnamefont {Eric}\ \bibnamefont
  {Lauga}}\ and\ \bibinfo {author} {\bibfnamefont {Todd~M.}\ \bibnamefont
  {Squires}},\ }\bibfield  {title} {\enquote {\bibinfo {title} {Brownian motion
  near a partial-slip boundary: {A} local probe of the no-slip condition},}\
  }\href {\doibase 10.1063/1.2083748} {\bibfield  {journal} {\bibinfo
  {journal} {Physics of Fluids}\ }\textbf {\bibinfo {volume} {17}},\ \bibinfo
  {pages} {103102} (\bibinfo {year} {2005})}\BibitemShut {NoStop}%
\bibitem [{\citenamefont {Ketzetzi}\ \emph {et~al.}(2020)\citenamefont
  {Ketzetzi}, \citenamefont {{de Graaf}}, \citenamefont {Doherty},\ and\
  \citenamefont {Kraft}}]{ketzetziSlipLengthDependent2020}%
  \BibitemOpen
  \bibfield  {author} {\bibinfo {author} {\bibfnamefont {Stefania}\
  \bibnamefont {Ketzetzi}}, \bibinfo {author} {\bibfnamefont {Joost}\
  \bibnamefont {{de Graaf}}}, \bibinfo {author} {\bibfnamefont {Rachel~P.}\
  \bibnamefont {Doherty}}, \ and\ \bibinfo {author} {\bibfnamefont
  {Daniela~J.}\ \bibnamefont {Kraft}},\ }\bibfield  {title} {\enquote {\bibinfo
  {title} {Slip {{Length Dependent Propulsion Speed}} of {{Catalytic Colloidal
  Swimmers}} near {{Walls}}},}\ }\href {\doibase
  10.1103/PhysRevLett.124.048002} {\bibfield  {journal} {\bibinfo  {journal}
  {Physical Review Letters}\ }\textbf {\bibinfo {volume} {124}},\ \bibinfo
  {pages} {048002} (\bibinfo {year} {2020})}\BibitemShut {NoStop}%
\bibitem [{\citenamefont {Muldowney}\ and\ \citenamefont
  {Higdon}(1995)}]{muldowney1995spectral}%
  \BibitemOpen
  \bibfield  {author} {\bibinfo {author} {\bibfnamefont {G.~P.}\ \bibnamefont
  {Muldowney}}\ and\ \bibinfo {author} {\bibfnamefont {J.~J.~L.}\ \bibnamefont
  {Higdon}},\ }\bibfield  {title} {\enquote {\bibinfo {title} {A spectral
  boundary element approach to three-dimensional {S}tokes flow},}\ }\href
  {\doibase 10.1017/S0022112095003260} {\bibfield  {journal} {\bibinfo
  {journal} {J. Fluid Mech.}\ }\textbf {\bibinfo {volume} {298}},\ \bibinfo
  {pages} {167--192} (\bibinfo {year} {1995})}\BibitemShut {NoStop}%
\bibitem [{\citenamefont {Pozrikidis}(1992)}]{pozrikidis1992}%
  \BibitemOpen
  \bibfield  {author} {\bibinfo {author} {\bibfnamefont {C.}~\bibnamefont
  {Pozrikidis}},\ }\href {\doibase 10.1017/CBO9780511624124} {\emph {\bibinfo
  {title} {Boundary Integral and Singularity Methods for Linearized Viscous
  Flow}}}\ (\bibinfo  {publisher} {Cambridge University Press},\ \bibinfo
  {year} {1992})\BibitemShut {NoStop}%
\bibitem [{\citenamefont {Leal}(2007)}]{leal2007advanced}%
  \BibitemOpen
  \bibfield  {author} {\bibinfo {author} {\bibfnamefont {L.~G.}\ \bibnamefont
  {Leal}},\ }\href
  {https://www.cambridge.org/core/books/advanced-transport-phenomena/9F2A633638780413DA73C6CB70A1D341}
  {\emph {\bibinfo {title} {Advanced transport phenomena: {F}luid mechanics and
  convective transport processes}}}\ (\bibinfo  {publisher} {Cambridge
  University Press},\ \bibinfo {year} {2007})\BibitemShut {NoStop}%
\bibitem [{\citenamefont {Kim}\ and\ \citenamefont {Karrila}(1991)}]{kim2005}%
  \BibitemOpen
  \bibfield  {author} {\bibinfo {author} {\bibfnamefont {S.}~\bibnamefont
  {Kim}}\ and\ \bibinfo {author} {\bibfnamefont {S.~J.}\ \bibnamefont
  {Karrila}},\ }\href {\doibase 10.1002/cite.330641004} {\emph {\bibinfo
  {title} {Microhydrodynamics: Principles and Selected Applications}}}\
  (\bibinfo  {publisher} {Butterworth-Heinemann, Boston},\ \bibinfo {year}
  {1991})\BibitemShut {NoStop}%
\bibitem [{\citenamefont {Kim}(2015)}]{kim2015ellipsoidal}%
  \BibitemOpen
  \bibfield  {author} {\bibinfo {author} {\bibfnamefont {S.}~\bibnamefont
  {Kim}},\ }\bibfield  {title} {\enquote {\bibinfo {title} {Ellipsoidal
  microhydrodynamics without elliptic integrals and how to get there using
  linear operator theory},}\ }\href {\doibase 10.1021/acs.iecr.5b01552}
  {\bibfield  {journal} {\bibinfo  {journal} {Ind. Eng. Chem. Res.}\ }\textbf
  {\bibinfo {volume} {54}},\ \bibinfo {pages} {10497--10501} (\bibinfo {year}
  {2015})}\BibitemShut {NoStop}%
\bibitem [{\citenamefont {Mazur}\ and\ \citenamefont
  {Saarloos}(1982)}]{mazur1982}%
  \BibitemOpen
  \bibfield  {author} {\bibinfo {author} {\bibfnamefont {P.}~\bibnamefont
  {Mazur}}\ and\ \bibinfo {author} {\bibfnamefont {W.~van}\ \bibnamefont
  {Saarloos}},\ }\bibfield  {title} {\enquote {\bibinfo {title} {Many-sphere
  hydrodynamic interactions and mobilities in a suspension},}\ }\href {\doibase
  10.1016/0378-4371(82)90127-3} {\bibfield  {journal} {\bibinfo  {journal}
  {Physica A: Stat. Mech. Appl.}\ }\textbf {\bibinfo {volume} {115}},\ \bibinfo
  {pages} {21--57} (\bibinfo {year} {1982})}\BibitemShut {NoStop}%
\bibitem [{\citenamefont {Singh}(2018)}]{singh2018microhydrodynamics}%
  \BibitemOpen
  \bibfield  {author} {\bibinfo {author} {\bibfnamefont {R.}~\bibnamefont
  {Singh}},\ }\emph {\bibinfo {title} {Microhydrodynamics of active colloids
  [HBNI Th131]}},\ \href {https://www.imsc.res.in/xmlui/handle/123456789/418}
  {Ph.D. thesis},\ \bibinfo  {school} {HBNI} (\bibinfo {year}
  {2018})\BibitemShut {NoStop}%
\bibitem [{\citenamefont {Ishikawa}\ \emph {et~al.}(2006)\citenamefont
  {Ishikawa}, \citenamefont {Simmonds},\ and\ \citenamefont
  {Pedley}}]{ishikawa2006}%
  \BibitemOpen
  \bibfield  {author} {\bibinfo {author} {\bibfnamefont {T.}~\bibnamefont
  {Ishikawa}}, \bibinfo {author} {\bibfnamefont {M.~P.}\ \bibnamefont
  {Simmonds}}, \ and\ \bibinfo {author} {\bibfnamefont {T.~J.}\ \bibnamefont
  {Pedley}},\ }\bibfield  {title} {\enquote {\bibinfo {title} {Hydrodynamic
  interaction of two swimming model micro-organisms},}\ }\href {\doibase
  10.1017/S0022112006002631} {\bibfield  {journal} {\bibinfo  {journal} {J.
  Fluid Mech.}\ }\textbf {\bibinfo {volume} {568}},\ \bibinfo {pages}
  {119--160} (\bibinfo {year} {2006})}\BibitemShut {NoStop}%
\bibitem [{\citenamefont {Swan}\ \emph {et~al.}(2011)\citenamefont {Swan},
  \citenamefont {Brady}, \citenamefont {Moore},\ and\ \citenamefont {{ChE
  174}}}]{swanModelingHydrodynamicSelfpropulsion2011}%
  \BibitemOpen
  \bibfield  {author} {\bibinfo {author} {\bibfnamefont {James~W.}\
  \bibnamefont {Swan}}, \bibinfo {author} {\bibfnamefont {John~F.}\
  \bibnamefont {Brady}}, \bibinfo {author} {\bibfnamefont {Rachel~S.}\
  \bibnamefont {Moore}}, \ and\ \bibinfo {author} {\bibnamefont {{ChE 174}}},\
  }\bibfield  {title} {\enquote {\bibinfo {title} {Modeling hydrodynamic
  self-propulsion with {Stokesian} {Dynamics}. {Or} teaching {Stokesian}
  {Dynamics} to swim},}\ }\href {\doibase 10.1063/1.3594790} {\bibfield
  {journal} {\bibinfo  {journal} {Physics of Fluids}\ }\textbf {\bibinfo
  {volume} {23}},\ \bibinfo {pages} {071901} (\bibinfo {year}
  {2011})}\BibitemShut {NoStop}%
\bibitem [{\citenamefont {Gradshteyn}\ and\ \citenamefont
  {Ryzhik}(2014)}]{gradshteyn2014table}%
  \BibitemOpen
  \bibfield  {author} {\bibinfo {author} {\bibfnamefont {I.~S.}\ \bibnamefont
  {Gradshteyn}}\ and\ \bibinfo {author} {\bibfnamefont {I.~M.}\ \bibnamefont
  {Ryzhik}},\ }\href@noop {} {\emph {\bibinfo {title} {Table of integrals,
  series, and products}}}\ (\bibinfo  {publisher} {Academic press},\ \bibinfo
  {year} {2014})\BibitemShut {NoStop}%
\bibitem [{\citenamefont {Anderson}\ and\ \citenamefont
  {Prieve}(1991)}]{anderson1991}%
  \BibitemOpen
  \bibfield  {author} {\bibinfo {author} {\bibfnamefont {J.~L.}\ \bibnamefont
  {Anderson}}\ and\ \bibinfo {author} {\bibfnamefont {D.~C.}\ \bibnamefont
  {Prieve}},\ }\bibfield  {title} {\enquote {\bibinfo {title} {Diffusiophoresis
  caused by gradients of strongly adsorbing solutes},}\ }\href {\doibase
  10.1021/la00050a035} {\bibfield  {journal} {\bibinfo  {journal} {Langmuir}\
  }\textbf {\bibinfo {volume} {7}},\ \bibinfo {pages} {403--406} (\bibinfo
  {year} {1991})}\BibitemShut {NoStop}%
\bibitem [{\citenamefont {Stone}\ and\ \citenamefont
  {Samuel}(1996)}]{stone1996}%
  \BibitemOpen
  \bibfield  {author} {\bibinfo {author} {\bibfnamefont {H.~A.}\ \bibnamefont
  {Stone}}\ and\ \bibinfo {author} {\bibfnamefont {A.~D.~T.}\ \bibnamefont
  {Samuel}},\ }\bibfield  {title} {\enquote {\bibinfo {title} {Propulsion of
  microorganisms by surface distortions},}\ }\href {\doibase
  10.1103/PhysRevLett.77.4102} {\bibfield  {journal} {\bibinfo  {journal}
  {Phys. Rev. Lett.}\ }\textbf {\bibinfo {volume} {77}},\ \bibinfo {pages}
  {4102--4104} (\bibinfo {year} {1996})}\BibitemShut {NoStop}%
\bibitem [{\citenamefont {Batchelor}(1970)}]{batchelor1970stress}%
  \BibitemOpen
  \bibfield  {author} {\bibinfo {author} {\bibfnamefont {G.~K.}\ \bibnamefont
  {Batchelor}},\ }\bibfield  {title} {\enquote {\bibinfo {title} {The stress
  system in a suspension of force-free particles},}\ }\href {\doibase
  10.1017/S0022112070000745} {\bibfield  {journal} {\bibinfo  {journal} {J.
  Fluid Mech.}\ }\textbf {\bibinfo {volume} {41}},\ \bibinfo {pages} {545--570}
  (\bibinfo {year} {1970})}\BibitemShut {NoStop}%
\bibitem [{\citenamefont {Lauga}\ and\ \citenamefont
  {Michelin}(2016)}]{laugaStressletsInducedActive2016}%
  \BibitemOpen
  \bibfield  {author} {\bibinfo {author} {\bibfnamefont {Eric}\ \bibnamefont
  {Lauga}}\ and\ \bibinfo {author} {\bibfnamefont {S{\'e}bastien}\ \bibnamefont
  {Michelin}},\ }\bibfield  {title} {\enquote {\bibinfo {title} {Stresslets
  {Induced} by {Active} {Swimmers}},}\ }\href {\doibase
  10.1103/PhysRevLett.117.148001} {\bibfield  {journal} {\bibinfo  {journal}
  {Physical Review Letters}\ }\textbf {\bibinfo {volume} {117}},\ \bibinfo
  {pages} {148001} (\bibinfo {year} {2016})}\BibitemShut {NoStop}%
\bibitem [{\citenamefont {Nasouri}\ and\ \citenamefont
  {Elfring}(2018)}]{nasouriHigherorderForceMoments2018}%
  \BibitemOpen
  \bibfield  {author} {\bibinfo {author} {\bibfnamefont {Babak}\ \bibnamefont
  {Nasouri}}\ and\ \bibinfo {author} {\bibfnamefont {Gwynn~J.}\ \bibnamefont
  {Elfring}},\ }\bibfield  {title} {\enquote {\bibinfo {title} {Higher-order
  force moments of active particles},}\ }\href {\doibase
  10.1103/PhysRevFluids.3.044101} {\bibfield  {journal} {\bibinfo  {journal}
  {Physical Review Fluids}\ }\textbf {\bibinfo {volume} {3}},\ \bibinfo {pages}
  {044101} (\bibinfo {year} {2018})},\ \bibinfo {note} {publisher: American
  Physical Society}\BibitemShut {NoStop}%
\bibitem [{\citenamefont {Hauge}\ and\ \citenamefont
  {Martin-L{\"o}f}(1973)}]{hauge1973fluctuating}%
  \BibitemOpen
  \bibfield  {author} {\bibinfo {author} {\bibfnamefont {E.~H.}\ \bibnamefont
  {Hauge}}\ and\ \bibinfo {author} {\bibfnamefont {A.}~\bibnamefont
  {Martin-L{\"o}f}},\ }\bibfield  {title} {\enquote {\bibinfo {title}
  {Fluctuating hydrodynamics and {B}rownian motion},}\ }\href {\doibase
  10.1007/BF01030307} {\bibfield  {journal} {\bibinfo  {journal} {J. Stat.
  Phys.}\ }\textbf {\bibinfo {volume} {7}},\ \bibinfo {pages} {259--281}
  (\bibinfo {year} {1973})}\BibitemShut {NoStop}%
\bibitem [{\citenamefont {Fox}\ and\ \citenamefont
  {Uhlenbeck}(1970)}]{fox1970contributions}%
  \BibitemOpen
  \bibfield  {author} {\bibinfo {author} {\bibfnamefont {R.~F.}\ \bibnamefont
  {Fox}}\ and\ \bibinfo {author} {\bibfnamefont {G.~E.}\ \bibnamefont
  {Uhlenbeck}},\ }\bibfield  {title} {\enquote {\bibinfo {title} {Contributions
  to non-equilibrium thermodynamics. {I}. {T}heory of hydrodynamical
  fluctuations},}\ }\href {\doibase 10.1063/1.1693183} {\bibfield  {journal}
  {\bibinfo  {journal} {Phys. Fluids}\ }\textbf {\bibinfo {volume} {13}},\
  \bibinfo {pages} {1893--1902} (\bibinfo {year} {1970})}\BibitemShut {NoStop}%
\bibitem [{\citenamefont {Bedeaux}\ and\ \citenamefont
  {Mazur}(1974)}]{bedeaux1974brownian}%
  \BibitemOpen
  \bibfield  {author} {\bibinfo {author} {\bibfnamefont {D.}~\bibnamefont
  {Bedeaux}}\ and\ \bibinfo {author} {\bibfnamefont {P.}~\bibnamefont
  {Mazur}},\ }\bibfield  {title} {\enquote {\bibinfo {title} {Brownian motion
  and fluctuating hydrodynamics},}\ }\href {\doibase
  10.1016/0031-8914(74)90198-0} {\bibfield  {journal} {\bibinfo  {journal}
  {Physica}\ }\textbf {\bibinfo {volume} {76}},\ \bibinfo {pages} {247--258}
  (\bibinfo {year} {1974})}\BibitemShut {NoStop}%
\bibitem [{\citenamefont {Roux}(1992)}]{roux1992brownian}%
  \BibitemOpen
  \bibfield  {author} {\bibinfo {author} {\bibfnamefont {J.-N.}\ \bibnamefont
  {Roux}},\ }\bibfield  {title} {\enquote {\bibinfo {title} {Brownian particles
  at different times scales: a new derivation of the {S}moluchowski
  equation},}\ }\href {\doibase 10.1016/0378-4371(92)90330-S} {\bibfield
  {journal} {\bibinfo  {journal} {Physica A: Stat. Mech. Appl.}\ }\textbf
  {\bibinfo {volume} {188}},\ \bibinfo {pages} {526--552} (\bibinfo {year}
  {1992})}\BibitemShut {NoStop}%
\bibitem [{\citenamefont {Zwanzig}(1964)}]{zwanzig1964hydrodynamic}%
  \BibitemOpen
  \bibfield  {author} {\bibinfo {author} {\bibfnamefont {R.}~\bibnamefont
  {Zwanzig}},\ }\bibfield  {title} {\enquote {\bibinfo {title} {Hydrodynamic
  fluctuations and {S}tokes law friction},}\ }\href@noop {} {\bibfield
  {journal} {\bibinfo  {journal} {J. Res. Natl. Bur. Std.(US)}\ }\textbf
  {\bibinfo {volume} {68B}},\ \bibinfo {pages} {143--145} (\bibinfo {year}
  {1964})}\BibitemShut {NoStop}%
\bibitem [{\citenamefont {Chow}(1973)}]{chow_simultaneous_1973}%
  \BibitemOpen
  \bibfield  {author} {\bibinfo {author} {\bibfnamefont {T.~S.}\ \bibnamefont
  {Chow}},\ }\bibfield  {title} {\enquote {\bibinfo {title} {Simultaneous
  translational and rotational {Brownian} movement of particles of arbitrary
  shape},}\ }\href {\doibase 10.1063/1.1694169} {\bibfield  {journal} {\bibinfo
   {journal} {The Physics of Fluids}\ }\textbf {\bibinfo {volume} {16}},\
  \bibinfo {pages} {31--34} (\bibinfo {year} {1973})},\ \bibinfo {note}
  {publisher: American Institute of Physics}\BibitemShut {NoStop}%
\bibitem [{\citenamefont {Cichocki}\ \emph {et~al.}(1994)\citenamefont
  {Cichocki}, \citenamefont {Felderhof}, \citenamefont {Hinsen}, \citenamefont
  {Wajnryb},\ and\ \citenamefont {Blawzdziewicz}}]{cichocki1994friction}%
  \BibitemOpen
  \bibfield  {author} {\bibinfo {author} {\bibfnamefont {B.}~\bibnamefont
  {Cichocki}}, \bibinfo {author} {\bibfnamefont {B.~U.}\ \bibnamefont
  {Felderhof}}, \bibinfo {author} {\bibfnamefont {K.}~\bibnamefont {Hinsen}},
  \bibinfo {author} {\bibfnamefont {E.}~\bibnamefont {Wajnryb}}, \ and\
  \bibinfo {author} {\bibfnamefont {J.}~\bibnamefont {Blawzdziewicz}},\
  }\bibfield  {title} {\enquote {\bibinfo {title} {Friction and mobility of
  many spheres in {S}tokes flow},}\ }\href {\doibase 10.1063/1.466366}
  {\bibfield  {journal} {\bibinfo  {journal} {J. Chem. Phys.}\ }\textbf
  {\bibinfo {volume} {100}},\ \bibinfo {pages} {3780--3790} (\bibinfo {year}
  {1994})}\BibitemShut {NoStop}%
\bibitem [{\citenamefont {Cichocki}\ \emph {et~al.}(2000)\citenamefont
  {Cichocki}, \citenamefont {Jones}, \citenamefont {Kutteh},\ and\
  \citenamefont {Wajnryb}}]{cichocki_friction_2000}%
  \BibitemOpen
  \bibfield  {author} {\bibinfo {author} {\bibfnamefont {B.}~\bibnamefont
  {Cichocki}}, \bibinfo {author} {\bibfnamefont {R.~B.}\ \bibnamefont {Jones}},
  \bibinfo {author} {\bibfnamefont {Ramzi}\ \bibnamefont {Kutteh}}, \ and\
  \bibinfo {author} {\bibfnamefont {E.}~\bibnamefont {Wajnryb}},\ }\bibfield
  {title} {\enquote {\bibinfo {title} {Friction and mobility for colloidal
  spheres in {Stokes} flow near a boundary: {The} multipole method and
  applications},}\ }\href {\doibase 10.1063/1.480894} {\bibfield  {journal}
  {\bibinfo  {journal} {J. Chem. Phys.}\ }\textbf {\bibinfo {volume} {112}},\
  \bibinfo {pages} {2548--2561} (\bibinfo {year} {2000})}\BibitemShut {NoStop}%
\bibitem [{\citenamefont {Corona}\ \emph {et~al.}(2017)\citenamefont {Corona},
  \citenamefont {Greengard}, \citenamefont {Rachh},\ and\ \citenamefont
  {Veerapaneni}}]{corona2017integralequation}%
  \BibitemOpen
  \bibfield  {author} {\bibinfo {author} {\bibfnamefont {Eduardo}\ \bibnamefont
  {Corona}}, \bibinfo {author} {\bibfnamefont {Leslie}\ \bibnamefont
  {Greengard}}, \bibinfo {author} {\bibfnamefont {Manas}\ \bibnamefont
  {Rachh}}, \ and\ \bibinfo {author} {\bibfnamefont {Shravan}\ \bibnamefont
  {Veerapaneni}},\ }\bibfield  {title} {\enquote {\bibinfo {title} {An integral
  equation formulation for rigid bodies in stokes flow in three dimensions},}\
  }\href {\doibase https://doi.org/10.1016/j.jcp.2016.12.018} {\bibfield
  {journal} {\bibinfo  {journal} {Journal of Computational Physics}\ }\textbf
  {\bibinfo {volume} {332}},\ \bibinfo {pages} {504--519} (\bibinfo {year}
  {2017})}\BibitemShut {NoStop}%
\bibitem [{\citenamefont {Blake}(1971{\natexlab{b}})}]{blake1971c}%
  \BibitemOpen
  \bibfield  {author} {\bibinfo {author} {\bibfnamefont {J.~R.}\ \bibnamefont
  {Blake}},\ }\bibfield  {title} {\enquote {\bibinfo {title} {A note on the
  image system for a {S}tokeslet in a no-slip boundary},}\ }\href {\doibase
  10.1017/S0305004100049902} {\bibfield  {journal} {\bibinfo  {journal} {Proc.
  Camb. Phil. Soc.}\ }\textbf {\bibinfo {volume} {70}},\ \bibinfo {pages}
  {303--310} (\bibinfo {year} {1971}{\natexlab{b}})}\BibitemShut {NoStop}%
\bibitem [{\citenamefont {Aderogba}\ and\ \citenamefont
  {Blake}(1978)}]{aderogba1978action}%
  \BibitemOpen
  \bibfield  {author} {\bibinfo {author} {\bibfnamefont {K.}~\bibnamefont
  {Aderogba}}\ and\ \bibinfo {author} {\bibfnamefont {J.~R.}\ \bibnamefont
  {Blake}},\ }\bibfield  {title} {\enquote {\bibinfo {title} {Action of a force
  near the planar surface between semi-infinite immiscible liquids at very low
  {R}eynolds numbers},}\ }\href {\doibase 10.1017/S0004972700008819} {\bibfield
   {journal} {\bibinfo  {journal} {Bull. Australian Math. Soc.}\ }\textbf
  {\bibinfo {volume} {19}},\ \bibinfo {pages} {309--318} (\bibinfo {year}
  {1978})}\BibitemShut {NoStop}%
\bibitem [{\citenamefont {Lee}\ \emph {et~al.}(1979)\citenamefont {Lee},
  \citenamefont {Chadwick},\ and\ \citenamefont {Leal}}]{lee_motion_1979}%
  \BibitemOpen
  \bibfield  {author} {\bibinfo {author} {\bibfnamefont {S.~H.}\ \bibnamefont
  {Lee}}, \bibinfo {author} {\bibfnamefont {R.~S.}\ \bibnamefont {Chadwick}}, \
  and\ \bibinfo {author} {\bibfnamefont {L.~G.}\ \bibnamefont {Leal}},\
  }\bibfield  {title} {\enquote {\bibinfo {title} {Motion of a sphere in the
  presence of a plane interface. {Part} 1. {An} approximate solution by
  generalization of the method of {Lorentz}},}\ }\href {\doibase
  10.1017/S0022112079001981} {\bibfield  {journal} {\bibinfo  {journal} {J.
  Fluid Mech.}\ }\textbf {\bibinfo {volume} {93}},\ \bibinfo {pages} {705--726}
  (\bibinfo {year} {1979})},\ \bibinfo {note} {publisher: Cambridge University
  Press}\BibitemShut {NoStop}%
\bibitem [{\citenamefont {Yang}\ and\ \citenamefont
  {Leal}(1984)}]{yang_particle_1984}%
  \BibitemOpen
  \bibfield  {author} {\bibinfo {author} {\bibfnamefont {S.-M.}\ \bibnamefont
  {Yang}}\ and\ \bibinfo {author} {\bibfnamefont {L.~G.}\ \bibnamefont
  {Leal}},\ }\bibfield  {title} {\enquote {\bibinfo {title} {Particle motion in
  {Stokes} flow near a plane fluid–fluid interface. {Part} 2. {Linear} shear
  and axisymmetric straining flows},}\ }\href {\doibase
  10.1017/S0022112084002652} {\bibfield  {journal} {\bibinfo  {journal} {J.
  Fluid Mech.}\ }\textbf {\bibinfo {volume} {149}},\ \bibinfo {pages}
  {275--304} (\bibinfo {year} {1984})},\ \bibinfo {note} {publisher: Cambridge
  University Press}\BibitemShut {NoStop}%
\bibitem [{\citenamefont {Falade}(1986)}]{falade_hydrodynamic_1986}%
  \BibitemOpen
  \bibfield  {author} {\bibinfo {author} {\bibfnamefont {A.}~\bibnamefont
  {Falade}},\ }\bibfield  {title} {\enquote {\bibinfo {title} {Hydrodynamic
  resistance of an arbitrary particle translating and rotating near a fluid
  interface},}\ }\href {\doibase 10.1016/0301-9322(86)90053-4} {\bibfield
  {journal} {\bibinfo  {journal} {International Journal of Multiphase Flow}\
  }\textbf {\bibinfo {volume} {12}},\ \bibinfo {pages} {807--837} (\bibinfo
  {year} {1986})}\BibitemShut {NoStop}%
\bibitem [{\citenamefont {Swan}\ and\ \citenamefont
  {Brady}(2007)}]{swan2007simulation}%
  \BibitemOpen
  \bibfield  {author} {\bibinfo {author} {\bibfnamefont {J.~W.}\ \bibnamefont
  {Swan}}\ and\ \bibinfo {author} {\bibfnamefont {J.~F.}\ \bibnamefont
  {Brady}},\ }\bibfield  {title} {\enquote {\bibinfo {title} {Simulation of
  hydrodynamically interacting particles near a no-slip boundary},}\ }\href
  {\doibase 10.1063/1.2803837} {\bibfield  {journal} {\bibinfo  {journal}
  {Phys. Fluids}\ }\textbf {\bibinfo {volume} {19}},\ \bibinfo {pages} {113306}
  (\bibinfo {year} {2007})}\BibitemShut {NoStop}%
\bibitem [{\citenamefont {B{\l}awzdziewicz}\ \emph {et~al.}(2010)\citenamefont
  {B{\l}awzdziewicz}, \citenamefont {Ekiel-Jezewska},\ and\ \citenamefont
  {Wajnryb}}]{blawzdziewicz_motion_2010}%
  \BibitemOpen
  \bibfield  {author} {\bibinfo {author} {\bibfnamefont {J.}~\bibnamefont
  {B{\l}awzdziewicz}}, \bibinfo {author} {\bibfnamefont {M.~L.}\ \bibnamefont
  {Ekiel-Jezewska}}, \ and\ \bibinfo {author} {\bibfnamefont {E.}~\bibnamefont
  {Wajnryb}},\ }\bibfield  {title} {\enquote {\bibinfo {title} {Motion of a
  spherical particle near a planar fluid-fluid interface: {The} effect of
  surface incompressibility},}\ }\href {\doibase 10.1063/1.3475197} {\bibfield
  {journal} {\bibinfo  {journal} {J. Chem. Phys.}\ }\textbf {\bibinfo {volume}
  {133}},\ \bibinfo {pages} {114702} (\bibinfo {year} {2010})},\ \bibinfo
  {note} {publisher: American Institute of Physics}\BibitemShut {NoStop}%
\bibitem [{\citenamefont {Liu}\ and\ \citenamefont
  {Prosperetti}(2010)}]{liu_wall_2010}%
  \BibitemOpen
  \bibfield  {author} {\bibinfo {author} {\bibfnamefont {Qianlong}\
  \bibnamefont {Liu}}\ and\ \bibinfo {author} {\bibfnamefont {Andrea}\
  \bibnamefont {Prosperetti}},\ }\bibfield  {title} {\enquote {\bibinfo {title}
  {Wall effects on a rotating sphere},}\ }\href {\doibase
  10.1017/S002211201000128X} {\bibfield  {journal} {\bibinfo  {journal} {J.
  Fluid Mech.}\ }\textbf {\bibinfo {volume} {657}},\ \bibinfo {pages} {1--21}
  (\bibinfo {year} {2010})}\BibitemShut {NoStop}%
\bibitem [{\citenamefont {Ladd}(1988)}]{ladd1988}%
  \BibitemOpen
  \bibfield  {author} {\bibinfo {author} {\bibfnamefont {A.~J.~C.}\
  \bibnamefont {Ladd}},\ }\bibfield  {title} {\enquote {\bibinfo {title}
  {Hydrodynamic interactions in a suspension of spherical particles},}\ }\href
  {\doibase 10.1063/1.454658} {\bibfield  {journal} {\bibinfo  {journal} {J.
  Chem. Phys.}\ }\textbf {\bibinfo {volume} {88}},\ \bibinfo {pages}
  {5051--5063} (\bibinfo {year} {1988})}\BibitemShut {NoStop}%
\bibitem [{\citenamefont {Brady}\ and\ \citenamefont
  {Bossis}(1988)}]{brady1988stokesian}%
  \BibitemOpen
  \bibfield  {author} {\bibinfo {author} {\bibfnamefont {J.~F.}\ \bibnamefont
  {Brady}}\ and\ \bibinfo {author} {\bibfnamefont {G.}~\bibnamefont {Bossis}},\
  }\bibfield  {title} {\enquote {\bibinfo {title} {Stokesian dynamics},}\
  }\href@noop {} {\bibfield  {journal} {\bibinfo  {journal} {Annu. Rev. Fluid
  Mech.}\ }\textbf {\bibinfo {volume} {20}},\ \bibinfo {pages} {111--157}
  (\bibinfo {year} {1988})}\BibitemShut {NoStop}%
\bibitem [{\citenamefont {Ichiki}(2002)}]{ichiki2002improvement}%
  \BibitemOpen
  \bibfield  {author} {\bibinfo {author} {\bibfnamefont {K.}~\bibnamefont
  {Ichiki}},\ }\bibfield  {title} {\enquote {\bibinfo {title} {Improvement of
  the {S}tokesian dynamics method for systems with a finite number of
  particles},}\ }\href {\doibase 10.1017/S0022112001006735} {\bibfield
  {journal} {\bibinfo  {journal} {J. Fluid Mech.}\ }\textbf {\bibinfo {volume}
  {452}},\ \bibinfo {pages} {231--262} (\bibinfo {year} {2002})}\BibitemShut
  {NoStop}%
\bibitem [{\citenamefont {Fiore}\ and\ \citenamefont
  {Swan}(2018)}]{fioreRapidSamplingStochastic2018}%
  \BibitemOpen
  \bibfield  {author} {\bibinfo {author} {\bibfnamefont {Andrew~M.}\
  \bibnamefont {Fiore}}\ and\ \bibinfo {author} {\bibfnamefont {James~W.}\
  \bibnamefont {Swan}},\ }\bibfield  {title} {\enquote {\bibinfo {title} {Rapid
  sampling of stochastic displacements in {Brownian} dynamics simulations with
  stresslet constraints},}\ }\href {\doibase 10.1063/1.5005887} {\bibfield
  {journal} {\bibinfo  {journal} {J. Chem. Phys.}\ }\textbf {\bibinfo {volume}
  {148}},\ \bibinfo {pages} {044114} (\bibinfo {year} {2018})},\ \bibinfo
  {note} {publisher: American Institute of Physics}\BibitemShut {NoStop}%
\bibitem [{\citenamefont {Ishikawa}\ and\ \citenamefont
  {Pedley}(2007)}]{ishikawa2007rheology}%
  \BibitemOpen
  \bibfield  {author} {\bibinfo {author} {\bibfnamefont {T.}~\bibnamefont
  {Ishikawa}}\ and\ \bibinfo {author} {\bibfnamefont {T.~J.}\ \bibnamefont
  {Pedley}},\ }\bibfield  {title} {\enquote {\bibinfo {title} {The rheology of
  a semi-dilute suspension of swimming model micro-organisms},}\ }\href
  {\doibase 10.1017/S0022112007007835} {\bibfield  {journal} {\bibinfo
  {journal} {J. Fluid Mech}\ }\textbf {\bibinfo {volume} {588}},\ \bibinfo
  {pages} {399--435} (\bibinfo {year} {2007})}\BibitemShut {NoStop}%
\bibitem [{\citenamefont {Damour}\ and\ \citenamefont
  {Iyer}(1991)}]{damour_multipole_1991}%
  \BibitemOpen
  \bibfield  {author} {\bibinfo {author} {\bibfnamefont {T.}~\bibnamefont
  {Damour}}\ and\ \bibinfo {author} {\bibfnamefont {Bala~R.}\ \bibnamefont
  {Iyer}},\ }\bibfield  {title} {\enquote {\bibinfo {title} {Multipole analysis
  for electromagnetism and linearized gravity with irreducible cartesian
  tensors},}\ }\href {\doibase 10.1103/PhysRevD.43.3259} {\bibfield  {journal}
  {\bibinfo  {journal} {Phys. Rev. D}\ }\textbf {\bibinfo {volume} {43}},\
  \bibinfo {pages} {3259--3272} (\bibinfo {year} {1991})}\BibitemShut {NoStop}%
\bibitem [{\citenamefont
  {Applequist}(2002)}]{applequist_maxwellcartesian_2002}%
  \BibitemOpen
  \bibfield  {author} {\bibinfo {author} {\bibfnamefont {Jon}\ \bibnamefont
  {Applequist}},\ }\bibfield  {title} {\enquote {\bibinfo {title}
  {Maxwell-{Cartesian} spherical harmonics in multipole potentials and atomic
  orbitals},}\ }\href {\doibase 10.1007/s00214-001-0301-2} {\bibfield
  {journal} {\bibinfo  {journal} {Theoretical Chemistry Accounts}\ }\textbf
  {\bibinfo {volume} {107}},\ \bibinfo {pages} {103--115} (\bibinfo {year}
  {2002})}\BibitemShut {NoStop}%
\bibitem [{\citenamefont {Greengard}\ and\ \citenamefont
  {Rokhlin}(1987)}]{greengard1987fast}%
  \BibitemOpen
  \bibfield  {author} {\bibinfo {author} {\bibfnamefont {L.}~\bibnamefont
  {Greengard}}\ and\ \bibinfo {author} {\bibfnamefont {Vladimir}\ \bibnamefont
  {Rokhlin}},\ }\bibfield  {title} {\enquote {\bibinfo {title} {A fast
  algorithm for particle simulations},}\ }\href {\doibase
  10.1006/jcph.1997.5706} {\bibfield  {journal} {\bibinfo  {journal} {J. Comp.
  Phys.}\ }\textbf {\bibinfo {volume} {73}},\ \bibinfo {pages} {325--348}
  (\bibinfo {year} {1987})}\BibitemShut {NoStop}%
\bibitem [{\citenamefont {White}\ and\ \citenamefont
  {Head-Gordon}(1996)}]{white_rotating_1996}%
  \BibitemOpen
  \bibfield  {author} {\bibinfo {author} {\bibfnamefont {Christopher~A.}\
  \bibnamefont {White}}\ and\ \bibinfo {author} {\bibfnamefont {Martin}\
  \bibnamefont {Head-Gordon}},\ }\bibfield  {title} {\enquote {\bibinfo {title}
  {Rotating around the quartic angular momentum barrier in fast multipole
  method calculations},}\ }\href {\doibase 10.1063/1.472369} {\bibfield
  {journal} {\bibinfo  {journal} {The Journal of Chemical Physics}\ }\textbf
  {\bibinfo {volume} {105}},\ \bibinfo {pages} {5061--5067} (\bibinfo {year}
  {1996})},\ \bibinfo {note} {publisher: American Institute of
  Physics}\BibitemShut {NoStop}%
\bibitem [{\citenamefont {Dachsel}(2006)}]{dachsel_fast_2006}%
  \BibitemOpen
  \bibfield  {author} {\bibinfo {author} {\bibfnamefont {Holger}\ \bibnamefont
  {Dachsel}},\ }\bibfield  {title} {\enquote {\bibinfo {title} {Fast and
  accurate determination of the {Wigner} rotation matrices in the fast
  multipole method},}\ }\href {\doibase 10.1063/1.2194548} {\bibfield
  {journal} {\bibinfo  {journal} {The Journal of Chemical Physics}\ }\textbf
  {\bibinfo {volume} {124}},\ \bibinfo {pages} {144115} (\bibinfo {year}
  {2006})},\ \bibinfo {note} {publisher: American Institute of
  Physics}\BibitemShut {NoStop}%
\bibitem [{\citenamefont {Shanker}\ and\ \citenamefont
  {Huang}(2007)}]{shanker2007accelerated}%
  \BibitemOpen
  \bibfield  {author} {\bibinfo {author} {\bibfnamefont {B.}~\bibnamefont
  {Shanker}}\ and\ \bibinfo {author} {\bibfnamefont {H.}~\bibnamefont
  {Huang}},\ }\bibfield  {title} {\enquote {\bibinfo {title} {Accelerated
  {C}artesian expansions--a fast method for computing of potentials of the form
  {R}$^{-\nu}$ for all real $\nu$},}\ }\href {\doibase
  10.1016/j.jcp.2007.04.033} {\bibfield  {journal} {\bibinfo  {journal} {J.
  Comput. Phys.}\ }\textbf {\bibinfo {volume} {226}},\ \bibinfo {pages}
  {732--753} (\bibinfo {year} {2007})}\BibitemShut {NoStop}%
\bibitem [{\citenamefont {Power}\ and\ \citenamefont
  {Miranda}(1987)}]{power1987second}%
  \BibitemOpen
  \bibfield  {author} {\bibinfo {author} {\bibfnamefont {H.}~\bibnamefont
  {Power}}\ and\ \bibinfo {author} {\bibfnamefont {G.}~\bibnamefont
  {Miranda}},\ }\bibfield  {title} {\enquote {\bibinfo {title} {Second kind
  integral equation formulation of {S}tokes flows past a particle of arbitrary
  shape},}\ }\href {\doibase 10.1137/0147047} {\bibfield  {journal} {\bibinfo
  {journal} {SIAM J. Appl. Math.}\ }\textbf {\bibinfo {volume} {47}},\ \bibinfo
  {pages} {689--698} (\bibinfo {year} {1987})}\BibitemShut {NoStop}%
\bibitem [{\citenamefont {Crameri}(2021)}]{crameri_scientific_2021}%
  \BibitemOpen
  \bibfield  {author} {\bibinfo {author} {\bibfnamefont {Fabio}\ \bibnamefont
  {Crameri}},\ }\href {\doibase 10.5281/zenodo.5501399} {\enquote {\bibinfo
  {title} {Scientific colour maps},}\ } (\bibinfo {year} {2021})\BibitemShut
  {NoStop}%
\bibitem [{\citenamefont {Crameri}\ \emph {et~al.}(2020)\citenamefont
  {Crameri}, \citenamefont {Shephard},\ and\ \citenamefont
  {Heron}}]{crameri_misuse_2020}%
  \BibitemOpen
  \bibfield  {author} {\bibinfo {author} {\bibfnamefont {Fabio}\ \bibnamefont
  {Crameri}}, \bibinfo {author} {\bibfnamefont {Grace~E.}\ \bibnamefont
  {Shephard}}, \ and\ \bibinfo {author} {\bibfnamefont {Philip~J.}\
  \bibnamefont {Heron}},\ }\bibfield  {title} {\enquote {\bibinfo {title} {The
  misuse of colour in science communication},}\ }\href {\doibase
  10.1038/s41467-020-19160-7} {\bibfield  {journal} {\bibinfo  {journal}
  {Nature Communications}\ }\textbf {\bibinfo {volume} {11}},\ \bibinfo {pages}
  {5444} (\bibinfo {year} {2020})},\ \bibinfo {note} {number: 1 Publisher:
  Nature Publishing Group}\BibitemShut {NoStop}%
\bibitem [{\citenamefont {Daddi-Moussa-Ider}\ \emph {et~al.}(2021)\citenamefont
  {Daddi-Moussa-Ider}, \citenamefont {Nasouri}, \citenamefont {Vilfan},\ and\
  \citenamefont {Golestanian}}]{daddi-moussa-ider_optimal_2021}%
  \BibitemOpen
  \bibfield  {author} {\bibinfo {author} {\bibfnamefont {Abdallah}\
  \bibnamefont {Daddi-Moussa-Ider}}, \bibinfo {author} {\bibfnamefont {Babak}\
  \bibnamefont {Nasouri}}, \bibinfo {author} {\bibfnamefont {Andrej}\
  \bibnamefont {Vilfan}}, \ and\ \bibinfo {author} {\bibfnamefont {Ramin}\
  \bibnamefont {Golestanian}},\ }\bibfield  {title} {\enquote {\bibinfo {title}
  {Optimal swimmers can be pullers, pushers or neutral depending on the
  shape},}\ }\href {\doibase 10.1017/jfm.2021.562} {\bibfield  {journal}
  {\bibinfo  {journal} {Journal of Fluid Mechanics}\ }\textbf {\bibinfo
  {volume} {922}},\ \bibinfo {pages} {R5} (\bibinfo {year} {2021})}\BibitemShut
  {NoStop}%
\bibitem [{\citenamefont {Guo}\ \emph {et~al.}(2021)\citenamefont {Guo},
  \citenamefont {Zhu}, \citenamefont {Liu}, \citenamefont {Bonnet},\ and\
  \citenamefont {Veerapaneni}}]{guo_optimal_2021}%
  \BibitemOpen
  \bibfield  {author} {\bibinfo {author} {\bibfnamefont {Hanliang}\
  \bibnamefont {Guo}}, \bibinfo {author} {\bibfnamefont {Hai}\ \bibnamefont
  {Zhu}}, \bibinfo {author} {\bibfnamefont {Ruowen}\ \bibnamefont {Liu}},
  \bibinfo {author} {\bibfnamefont {Marc}\ \bibnamefont {Bonnet}}, \ and\
  \bibinfo {author} {\bibfnamefont {Shravan}\ \bibnamefont {Veerapaneni}},\
  }\bibfield  {title} {\enquote {\bibinfo {title} {Optimal slip velocities of
  micro-swimmers with arbitrary axisymmetric shapes},}\ }\href
  {https://www.cambridge.org/core/journals/journal-of-fluid-mechanics/article/optimal-slip-velocities-of-microswimmers-with-arbitrary-axisymmetric-shapes/52FD80312DEED50DC2E484ECC78BD674}
  {\bibfield  {journal} {\bibinfo  {journal} {Journal of Fluid Mechanics}\
  }\textbf {\bibinfo {volume} {910}},\ \bibinfo {pages} {A26} (\bibinfo {year}
  {2021})}\BibitemShut {NoStop}%
\end{thebibliography}
%

\end{document}